\documentclass[aps,prl,twocolumn,amsmath,amssymb,notitlepage,superscriptaddress,10pt]{revtex4-2}

\usepackage{amsmath}
\usepackage{amssymb}
\usepackage{amsbsy}
\usepackage{graphicx}
\usepackage{color}
\usepackage{layout}
\usepackage{printlen}

\usepackage{mathrsfs}
\usepackage{enumerate}
\usepackage{mathtools}
\usepackage[%
  colorlinks=true,
  urlcolor=blue,
  linkcolor=blue,
  citecolor=blue
]{hyperref}

\usepackage{tabularx,booktabs}

\DeclareMathOperator{\atan}{atan}
\DeclareMathOperator{\erf}{erf}
\DeclareMathOperator{\erfc}{erfc}

\newcommand\vvec{\underline}

\newcommand{\mcA}{\mathcal{A}}
\newcommand{\mcB}{\mathcal{B}}
\newcommand{\mcC}{\mathcal{C}}

\def \equi#1{\mathrel{\mathop{\kern 0pt\sim}\limits_{#1}}}

\begin{document}
    \title{Full stochastic dynamics of a tracer in a dense single-file system}
	
	\author{Alexis Poncet}
    \affiliation{CNRS, ENS de Lyon, LPENSL, UMR5672, 69342, Lyon cedex 07, France}
	
	\author{Aur\'elien Grabsch}
    \affiliation{Sorbonne Université, CNRS, Laboratoire de Physique Théorique de la
Matière Condensée (LPTMC), 4 Place Jussieu, 75005 Paris, France}
	
	\author{Olivier B\'enichou}
    \affiliation{Sorbonne Université, CNRS, Laboratoire de Physique Théorique de la
Matière Condensée (LPTMC), 4 Place Jussieu, 75005 Paris, France}

    \begin{abstract}
    Tracer diffusion in single-file systems, where particles are restricted to move on a line without passing each other, has been a fertile ground to investigate anomalous diffusion and strong memory effects.
    While the long-time behavior of such a tracer has been well studied, with a known subdiffusive dynamics and a Gaussian description for the rescaled position, the finer details of multi-time correlations remain poorly understood.
    This work focuses on the limit where almost all sites of a Symmetric Exclusion Process (SEP), a paradigmatic lattice model, are occupied. It extends beyond Gaussian descriptions and single-time statistics to address the multi-time correlation functions of the tracer in the SEP. 
    In this dense limit, we present a general relation between all $n$-time correlations of the non-Markovian tracer position process and the conditional probabilities of a single Markovian random walker.
    Using this relation, we derive explicit expressions for the four-time correlations and further explore important extensions: multiple tracers, non-equilibrium situations, and finite observation times.
    Our results underscore significant memory effects, strong temporal correlations, and the influence of initial conditions on long-time dynamics.
    \end{abstract}
	
	\maketitle
    
    \let\oldaddcontentsline\addcontentsline
    \renewcommand{\addcontentsline}[3]{}

    \textit{Introduction.---} 
The Symmetric Exclusion Process (SEP) and, in particular, the tracer diffusion within the SEP, are paradigmatic models in non-equilibrium statistical physics~\cite{Spohn1991,Mallick_2015}.
These models serve as minimal frameworks for understanding anomalous transport in crowded environments, such as porous media and biological systems~\cite{Hahn1996,Hofling2013}.
Due to their simplicity and rich underlying structure, they have attracted considerable interest from both mathematical and physical perspectives~\cite{Harris_1965,Arratia_1983,peligrad_2008,Illien2013,Hegde:2014,Krapivsky:2014,Krapivsky:2015a,Imamura_2017,Mallick2022,Grabsch2023,Grabsch2024}. 

It has been known for a long time that the strong constraints in the single-file geometry result in a subdiffusive behavior of the tracer at large time $\langle X(t)^2 \rangle\sim t^{1/2}$~\cite{Harris_1965,Arratia_1983}. In addition, the rescaled tracer position process converges to a Gaussian process in the long-time limit~\cite{Arratia_1983,peligrad_2008}. However, this Gaussian description captures only the leading-order statistical properties of the process and overlooks important finer details of the fluctuations. In recent years, a growing body of work has sought to go beyond Gaussian description. Notable advances have been made in understanding the deviation from this Gaussian behavior, encoded in higher order cumulants or large deviation functions~\cite{Illien2013,Hegde:2014,Krapivsky:2014,Krapivsky:2015a,Imamura_2017,Mallick2022,Grabsch2023,Grabsch2024}, and more recently the determination of the correlations between the tracer and the other particles~\cite{Poncet2021,Grabsch:2022,Grabsch2023}. 
These developments not only provided deeper insights into the mechanisms governing subdiffusion in single-file geometry, but also highlighted the central place of exclusion models inside the study of integrable systems~\cite{Grabsch2023,Mallick2022,Bettelheim2021,Krajenbrink2021,Krajenbrink2025}.

Despite this progress, essentially all existing results focus on single-time statistical properties, which do not provide a complete characterization of the tracer position process. The only result available in the literature concerns two-time correlations~\cite{peligrad_2008,Krapivsky_2015b,Sadhu_2015}. However, because of the interactions with the surrounding particles, the dynamics of the 
tracer are inherently non-Markovian and non-Gaussian, as evidenced by the non-vanishing higher-order cumulants. This makes the determination of all multi-time correlation functions of the position of the tracer a required step to fully characterize the tracer position process. Finally, in current state, the tracer position process in the SEP, even if fundamental, is not entirely characterized.

In this context, a key step to go further is the calculation of four-time correlation functions of the position of the tracer. Extending the analysis beyond Gaussian descriptions actually plays a role similar to early breakthroughs in the determination of higher-order cumulants at a single time, as achieved in recent years~\cite{Illien2013,Hegde:2014,Krapivsky:2014,Krapivsky:2015a,Imamura_2017,Mallick2022,Grabsch2023,Grabsch2024}. 

In this work, we investigate the stochastic process of a tracer in the SEP by focusing on the 
limit in which almost all sites are occupied. We reveal a strikingly simple and general relation between all $n$-time correlations of the non-Markovian tracer position process and conditional probabilities of a single Markovian random walker. Relying on this general relation, we derive explicit expressions for multi-time correlations up to the fourth order.
We extend these results in several important directions: (i) case of several tracers; (ii) case of non-equilibrium situations, which involve both an initial step of density and a biased tracer in a bath of unbiased particles; (iii) case of finite observation time, and not only long times.
In addition, by studying both annealed and quenched initial conditions, our results highlight the pronounced memory effects inherent to the SEP, the strong correlations between different time scales, and the crucial influence of initial conditions on the long-time dynamics~\cite{Leibovich_2013}.

    \begin{figure}
        \centering
        \includegraphics[scale=1]{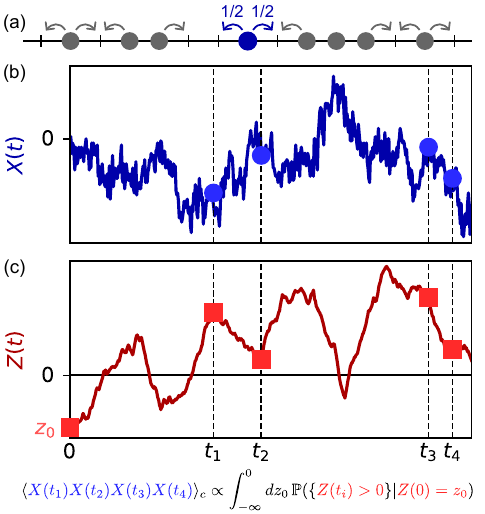} 
        \caption{
        \textbf{Tracer statistics from a Brownian motion.}
        (a) In the SEP, particles attempt to jump on neighboring sites with an exponential clock of unit rate. These symmetric jumps are rejected if the arrival site is occupied. We place a tracer (blue) at the origin and investigate its time evolution $X(t)$. (b) The fourth cumulant
        $\langle X(t_1)\dots X(t_4)\rangle_c = \langle X(t_1)\dots X(t_4)\rangle - \sum_{\{(ij),(kl)\}} \langle X(t_i)X(t_j)\rangle\langle X(t_k)X(t_l)\rangle$ does not vanish, in opposition with a Gaussian process. (c)
        In the dense limit, 
        this cumulant (as well as the higher-order ones) can be computed from conditional probabilities of a simple Brownian motion $Z(t)$, 
        see also Eqs.~\eqref{eq:kappa2nA}-\eqref{eq:PnP}.}
        \label{fig:sep}
    \end{figure}

\textit{Dense symmetric exclusion process.---} 
Our basic setup is a tracer particle in the SEP, which is initially at the origin, $X(t_0=0)=0$, see Fig.~\ref{fig:sep}a.
We aim at characterizing its time evolution
$X(t)$ in the dense limit, that is to say when the fraction $\rho$ of occupied sites (density) is $\rho\to 1$. 
The stochastic process is
fully defined by the knowledge of all the $n$-time correlations of $X$ at arbitrary times $0<t_1\leq \dots\leq t_n$, as illustrated in Fig.~\ref{fig:sep}b for $n=4$.
In turn, for a fixed initial configuration $\mathcal{C}$, these correlations are generated by the function
$G_n(\vvec\mu|\mathcal{C}) = \left\langle e^{\sum_i \mu_i X(t_i)} \right\rangle_\mathrm{evol}$ where $\vvec\mu=(\mu_1, \dots, \mu_n)$ and `$\mathrm{evol}$' denotes an average over the stochastic evolution of the system.
Importantly, there are two non-equivalent ways of averaging $G_n$ over the initial conditions, leading to two different cumulant-generating functions (CGFs) in the dense limit:
\begin{equation} \label{eq:psiAQ}
    \psi_n^A(\vvec\mu) = \lim_{\rho\to 1} \frac{\ln \langle  G_n\rangle_{\mathcal{C}}}{1-\rho},
    \quad
    \psi_n^Q(\vvec\mu) = \lim_{\rho\to 1} \frac{\langle \ln G_n\rangle_{\mathcal{C}}}{1-\rho}.
\end{equation}
The annealed CGF $\psi_n^A$ corresponds to a random initialization, while the quenched CGF $\psi_n^Q$ describes a ``typical'' initial configuration and is 
for instance obtained by placing the particles deterministically.
As is already known at the level of the covariance, Eqs.~\eqref{eq:kappa2Ares}-\eqref{eq:kappa2Qres} below, these two CGFs are different. 
This is a consequence of strong memory effects 
in the single-file geometry~\cite{Leibovich_2013,Poncet2021}.

In the following, we focus on the large time behavior of 
the $n$-time cumulants encoding the correlations of the stochastic process,
\begin{equation}
    \kappa_n^{A,Q}
    = \left.\frac{\partial^n\psi_n^{A,Q} 
    }{\partial\mu_1\dots\partial\mu_n}\right|_{\mu_i=0} = \lim_{\rho\to 1} \frac{\langle X(t_1) \dots X(t_n)\rangle_c^{A,Q}
    }{1-\rho}.
    \label{eq:kappa_n}
\end{equation}

    \textit{Computation of the CGFs.---} 
    We now give the main steps for the computation of the CGFs, Eq.~\eqref{eq:psiAQ}.
    As first noticed by Brummelhuis and Hilhorst~\cite{Brummelhuis_1989} for single-time observables in 2D systems, it is convenient to look at the dynamics of the vacancies, as in the high-density limit ($\rho\to 1$) they are independent~\cite{Illien2013,Poncet2018,Poncet2021,Poncet2022}.
    Here, we extend this approach to study multi-time observables~\footnote{Following Ref.~\cite{Poncet2022}, we also investigate the dynamics of the vacancies in continuous time and obtain results beyond the large time limit.}.

    We start from a finite system with $M$ vacancies, in which the generating function factorizes as
    $G_n(\vvec\mu|\mathcal{C}) = \prod_{j=1}^M G_n^{(1)}(\vvec\mu|\mathcal{Z}_j)$
    where $G_n^{(1)}$ is the analog of $G_n$ in a system with a \emph{single} vacancy at initial position $\mathcal{Z}_j$.
    Taking first the limit of infinite system size, then the limit of large time increments, Eq.~\eqref{eq:psiAQ} leads to the expressions for the multi-time CGFs
    $\psi_n^A(\vvec\mu) \sim \sqrt{2} \int_{-\infty}^\infty [\mathcal{P}_n(\vvec\mu, z)-1] dz$ 
    and $\psi_n^Q(\vvec\mu) \sim \sqrt{2} \int_{-\infty}^\infty \log\mathcal{P}_n(\vvec\mu, z) dz$ where
    $\mathcal{P}_n(\vvec\mu, z)$ is the large-time limit of $G_n^{(1)}(\vvec\mu|\mathcal{Z})$ with $z$ the continuous analog of the initial position $\mathcal{Z}$ of the single vacancy.
    Here and in the following, the symbol `$\sim$' denotes a limit of large time increments $t_i-t_{i-1}\to\infty$.
    To go further and obtain explicit expressions for $\mathcal{P}_n$, and thus for the CGFs, two ingredients are needed: 
    (i) At large time, a single vacancy behaves as a symmetric Brownian particle $Z$ of diffusion constant $1/2$, with propagator
    $K_\tau(z', z) = e^{-\frac{(z' - z)^2}{\tau}}/\sqrt{\pi\tau}$ which gives the probability density that $Z$ is at position $z'$ at time $t+\tau$ knowing that it was at $z$ at time $t$;
    (ii) In a single-vacancy system the tracer can move at most by one lattice spacing, and it does so when the vacancy crosses the origin. 
    We show that summing the contributions of the independent vacancies, these two ingredients eventually lead to the expression~\cite{SM}
    \begin{equation} \label{eq:PPn}
        \mathcal{P}_n(\vvec\mu, z_0) 
        = \int_{-\infty}^\infty \prod_{i=1}^n dz_i\,  
        K_{\tau_i}(z_i,z_{i-1}) e^{\frac{\nu_{i-1}-\nu_i}{2}\lambda_i},
    \end{equation}
    where $\nu_i = \mathrm{sign}(z_i)$ and $\lambda_i = \sum_{j=1}^i \mu_j$ are the generating parameters associated with the spatial increments $X(t_i)-X(t_{i-1})$. 
    
    \textit{Single-file cumulants.---} 
    The cumulants $\kappa_n$, defined in Eq.~\eqref{eq:kappa_n}, are conveniently obtained from the previous expressions.
    Since the motion of the tracer is symmetric, the cumulants of odd order vanish: $\kappa_{2n+1}^{A,Q}=0$. The annealed cumulants of even order can be readily expressed from Eqs.~\eqref{eq:kappa_n}-\eqref{eq:PPn}:
    \begin{align}
        \label{eq:kappa2nA}
      \kappa_{2n}^A &\sim 2\sqrt{2} \int_{-\infty}^0dz_0\, P_{2n}^+(z_0), \\
    \label{eq:PnP}
        P_n^+(z_0) &= \int_0^\infty \prod_{i=1}^n dz_i\,
        K_{\tau_i}(z_i,z_{i-1}).
    \end{align}
    We stress that $P_n^+$  is the probability that the random walker $Z$ defined above is at a positive position at all times $t_n$ knowing that it started from $z_0$ at time $t=0$, see Fig.~\ref{fig:sep}c. 
    Together, Eqs.~\eqref{eq:kappa2nA}-\eqref{eq:PnP} have a profound implication on the nature of the stochastic process realized by the tracer in the SEP: although strongly non Markovian, all its statistics can in fact be derived from the Markovian variable $Z$. The pronounced memory effects in the dense SEP are captured by a simple constrained random walk that remains positive.
    
    \begin{figure}
        \centering
        \includegraphics[width=\columnwidth]{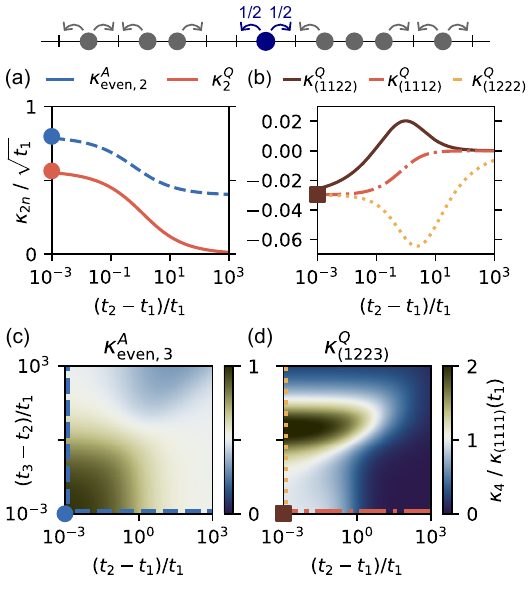} 
        \caption{
        \textbf{Unbiased tracer.}
        (a) 
        Correlation $\kappa_2(t_1, t_2)$ rescaled by $\sqrt{t_1}$ as a function of the ratio of increments $(t_2-t_1)/t_1$, for annealed (dashed blue, Eq.~\eqref{eq:kappa2Ares}) and quenched initial conditions (red, Eq.~\eqref{eq:kappa2Qres}). The circles represent the coefficient of the single-time variance.
        (b) Quenched cumulants of order $4$ involving two times $t_1$ and $t_2$. The square is the coefficient of the single-time cumulant.
        (c) Annealed cumulant of order $4$ involving three times rescaled by the single-time cumulant $\kappa_{(1111)}^A(t_1)$, as a function of the ratio of time increments. The dashed lines correspond to the curve in (a).
        (d) Quenched cumulant of order $4$, $\kappa_{(1223)}^Q$, rescaled by the single-time cumulant. The cuts correspond to the curves in (b).}
        \label{fig:unbiased}
    \end{figure}
    
    Another striking consequence of Eq.~\eqref{eq:PnP} is that all annealed cumulants involving times in a given set, $t_i\in \{t_1', \dots, t_k'\}$,
    are equal and depend only on this set of times. Denoting them by $\kappa^A_{\mathrm{even}, k}(t_1', \dots, t_k')$, we have for instance: 
    $\kappa^A_4(t_1', t_2', t_2', t_2') = \kappa^A_2(t_1', t_2') \equiv \kappa_{\mathrm{even},2}^A(t_1', t_2')$. 
    
    There is no such equality for quenched cumulants, where
    the growing complexity of the expressions with the order of the cumulants comes from the non-linear, logarithmic, expression of the cumulant generating function 
    with $\mathcal{P}_n$ derived above. 
    Note however that the quenched cumulants can still be expressed in terms of the random walk probability $P_n^+$~\cite{SM}, Eq.~\eqref{eq:PnP}. For example, 
    \begin{equation} \label{eq:kappa2Q}
        \kappa_2^Q \sim \kappa_2^A - 2\sqrt{2} \int_{-\infty}^0dz\, P_1^+(z,t_1)P_1^+(z,t_2).
    \end{equation}
    
    \textit{Explicit cumulants.---} From the previous framework, we now obtain  explicit expressions for the $n$-time cumulants $\kappa_n$, see \cite{Poncet_TracerDenseSingleFile_2025_2} for the symbolic computations.
    First, covariances are  obtained from  Eqs.~\eqref{eq:kappa2nA}-\eqref{eq:kappa2Q} using known Gaussian integrals~\cite{Owen1980}, see Fig.~\ref{fig:unbiased}a,
    \begin{align} \label{eq:kappa2Ares}
        \kappa_{\mathrm{even},2}^A(t_1,t_2) &\sim \frac{1}{\sqrt{2\pi}}\left(\sqrt{t_1} + \sqrt{t_2} - \sqrt{|t_2-t_1|}\right), \\
        \label{eq:kappa2Qres}
        \kappa_{2}^Q(t_1, t_2) &\sim \frac{1}{\sqrt{2\pi}}\left(\sqrt{t_1+t_2} - \sqrt{|t_2-t_1|}\right).
    \end{align} 
    They correspond to the known result for single-file systems at any density~\cite{Krapivsky_2015b}.
    In the annealed case, Eq.~\eqref{eq:kappa2Ares}
    encodes a fractional Brownian motion with Hurst index $H=1/4$, which is the Gaussian limit of the process~\cite{peligrad_2008}. 
    We stress that Eq.~\eqref{eq:kappa2Ares} also holds for arbitrary two-time cumulants $\langle X(t_1)^p X(t_2)^q\rangle_c$ in the dense limit.
    Aside from the different coefficient for the variance ($t_1=t_2$), 
    a major qualitative difference between annealed and quenched initial conditions, that will resurface later,  is the behavior of the covariance $\kappa_2(t_1, t_2)$ at large $t_2$ (with $t_1$ fixed). In the quenched case, it decays to zero, while not in the annealed case (it converges to half of the value of the variance), see Fig.~\ref{fig:unbiased}a.
    
    Importantly, our framework allows us to go  beyond Gaussianity. 
    We obtain the explicit expressions of the order-4 cumulants $\kappa_{\mathrm{even}, 4}^A$ and $\kappa^Q_{(ijkl)} = \kappa^Q_4(t_i, t_j, t_k, t_l)$, 
    \begin{widetext}
    \begin{gather}
        \label{eq:kappaAeven4}
        \kappa_{\mathrm{even}, 4}^A \sim \frac{1}{4\sqrt{2\pi}}\bigg[(1+a_{123}+a_{124}+a_{134})\sqrt{t_1} + (1+a_{234})\sqrt{t_2} + (1+b_{0123}) \sqrt{t_3} + (1+b_{0124}+b_{0134}+b_{0234})\sqrt{t_4}   \\ -
        (1+a_{234})\sqrt{t_2-t_1} - \sqrt{t_3-t_1} - (1+b_{1234})\sqrt{t_4-t_1} - a_{012}\sqrt{t_3-t_2} - a_{012}\sqrt{t_4-t_2} - (a_{013}+a_{023}-a_{123})\sqrt{t_4-t_3}\bigg], \nonumber \\
        \kappa^Q_{(1112)}\sim\frac{3\sqrt{t_1}}{2\sqrt\pi}\left[3A\left(\frac{t_1}{2(t_1+3t_2)}\right)-A\left(\frac{3t_1}{2(t_2-t_1)}\right)\right] + \frac{\sqrt{t_1+t_2}}{2\sqrt{2\pi}}\left[8-9A\left(\frac{(t_1+t_2)(t_1+3t_2)}{t_2^2}\right)-6A\left(\frac{t_2-t_1}{3(t_1+t_2)}\right)\right],
        \label{eq:kappa1112Qres}
    \end{gather}
    \end{widetext}
where we have introduced the elementary function
        $A(u) = ({2}/{\pi})\arctan\sqrt{u}$,
    and denoted $a_{ijk}=A\left(\frac{t_j-t_i}{t_k-t_j}\right)$ and 
        $b_{ijkl}=A\left(\frac{(t_j-t_i)(t_l-t_k)}{(t_l-t_i)(t_k-t_j)}\right)$. Several comments are in order. (i) 
    As expected, the annealed cumulant $\kappa^A_\mathrm{even,4}$ reduces to $\kappa^A_\mathrm{even,2}$, Eq.~\eqref{eq:kappa2Ares}, whenever only two times are involved (for instance $t_2=t_3=t_4$), and to $\kappa^A_\mathrm{even,1}$ when all times are equal. (ii) When $t_1=t_2$, the quenched cumulant $\kappa^Q_{(1112)}$ goes to the already non-trival single-time expression $\kappa^Q_{(1111)}\sim [7-9A(8)]\sqrt{t/\pi}$~\cite{Poncet2021}. (iii)
    The full expression of the 4-time quenched cumulant $\kappa_4^Q$
    is a sum of the factors $\sqrt{t_j\pm t_i}$ ($0\leq i<j\leq 4$) with coefficients expressed in term of the function $A$, see End matter.
    (iv) In Fig.~\ref{fig:unbiased}b, we plot the order-4 quenched cumulants involving two times: 
    their variation is non-monotonous and
    they can even change sign.
    (v) In Fig.~\ref{fig:unbiased}c-d, we contrast the annealed and quenched order-4 cumulants involving three times:   the quenched cumulant decays to zero at large time increments (with fixed $t_1$) while the annealed one does not, which is reminiscent of the behavior of the covariance  above;  the annealed cumulant decays monotonously with increasing time increments while the quenched cumulant shows non-monotonicity.

    As we now show, our framework to determine $n$-time correlations in a dense SEP can be extended in several important directions. 
    
    \textit{Two tracers.---} Generalizing  Ref.~\cite{Poncet2018}, we compute the covariance between two tracers of positions $X(t)$ and $Y(t)$, with initial separation $Y(0)-X(0) = L$, 
    a situation typically at play in two-tracer microrheology experiments. In the annealed and quenched cases, we obtain~\cite{SM}
    \begin{align}
    \label{eq:2tracersA}
        \lim_{\rho\to 1} \frac{\langle X(t_1) Y(t_2)\rangle_A}{1-\rho} &\sim \frac{g_L(t_1) + g_L(t_2) - g_L(|t_1-t_2|)}{\sqrt{2\pi}}, \\
        \label{eq:2tracersQ}
        \lim_{\rho\to 1} \frac{\langle X(t_1) Y(t_2)\rangle_Q}{1-\rho} &\sim \frac{g_L(t_1+t_2)- g_L(|t_1-t_2|)}{\sqrt{2\pi}},
    \end{align}
    with $g_L(t) = \sqrt{t}[e^{-\ell^2} - \sqrt{\pi} \ell \erfc \ell]$, $\ell = L/\sqrt{2t}$. 
    Eqs~\eqref{eq:2tracersA}-\eqref{eq:2tracersQ} are in line with the Edwards-Wilkinson Gaussian theory respectively with equilibrium and flat initial conditions, see End matter. These equations provide the generalization of the covariance, Eqs.~\eqref{eq:kappa2Ares}-\eqref{eq:kappa2Qres}, to the case of a $L\neq0$  initial separation.

    \textit{Step initial condition.---} 
    A paradigmatic  non-equilibrium extension in the context of exclusion processes  is the case of an initial step density profile, $\rho_+$ on the right of the TP, $\rho_-$ on the left~\cite{Derrida_2009,Derrida_2009a,Imamura_2017}.  Rescaling now the  cumulants, see Eq.~\eqref{eq:kappa_n}, by the average density $\rho = (\rho_- + \rho_+)/2$  and  defining the step parameter $\sigma = (\rho_- - \rho_+)/[2(1-\rho)]$~\cite{Poncet2021}, Eq.~\eqref{eq:PPn} takes an additional  multiplicative factor $(1+\nu_0\sigma)$ which has neat consequences:
    (i) The even cumulants $\kappa_{\mathrm{even},k}^A$ and $\kappa_{2n}^Q$ are identical to the equilibrium SEP, see Eqs.~\eqref{eq:kappa2nA}-\eqref{eq:kappa1112Qres};
    (ii) The odd annealed cumulants involving $k$ times are equal and proportional to the even cumulants $\kappa_{\mathrm{odd}, k}^A = \sigma\kappa_{\mathrm{even}, k}^A$;
    (iii) The odd quenched cumulants, beyond the lowest one
    $\kappa_1^Q \sim \sigma\sqrt{2t/\pi}$, have non-trivial expressions, see End matter.
    \begin{figure}
        \centering
        \includegraphics[width=\columnwidth]{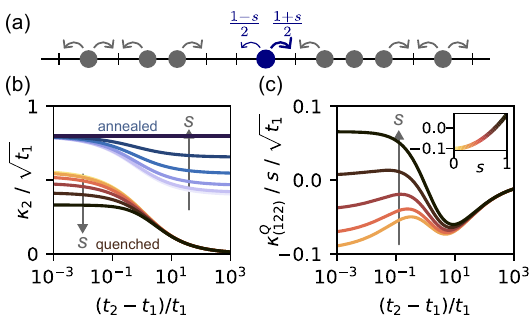}
        \caption{\textbf{Biased tracer.} (a) The jump probabilities of the tracer are no longer symmetric, with a bias $s$.
        (b) Covariance $\kappa_2$ for annealed (blue) and quenched (red) initial conditions for biases $s=0, 0.2, 0.4, 0.6, 0.8, 1$ (light to dark).
        (c) Scaled quenched cumulant $\kappa_{(122)}^Q/s$ with increasing bias (light to dark). The inset shows the value when $t_2=t_1$, that is to say the single-time cumulant $\kappa_{(111)}^Q/s$.}
        \label{fig:biased}
    \end{figure}

    \textit{Biased tracer.---} A second  important nonequilibrium extension, which has received attention both in the physics~\cite{Burlatsky_1996,Lizana2010,Illien2013,Poncet2021} and mathematics~\cite{Landim_1998} literature, is the case of a biased tracer particle in the SEP. The tracer has now  a probability 
    $(1 + s)/2$ to jump to the right and
    $p_- = (1-s)/2$ to jump to the left, see Fig.~\ref{fig:biased}a.  While Eq.~\eqref{eq:PPn} remains valid,  the Gaussian propagator $K$ is   replaced  by the biased kernel
    \begin{equation}
        K_{\tau, s}(z', z) = \frac{1}{\sqrt{\pi\tau}}\left[
        e^{-\frac{(z'-z)^2}{\tau}} - \nu' s e^{-\frac{(|z'|+|z|)^2}{\tau}}\right],
    \end{equation}
    where $\nu'$ is the sign of $z'$.
    Instead of the single quantity $P_n^+(z_0)$, Fig.~\ref{fig:sep}c, one now needs to compute the $2^{n-1}$ conditional probabilities associated with $Z(t_i)\gtrless 0$ for $i\geq 2$. Each probability is multiplied by a power of the bias $s$ linked with the number of time the random walker changes sign between successive times~\cite{SM}. For instance, when computing $\kappa_2^A$ with Eq.~\eqref{eq:kappa2nA}, $P_2^+(z_0)$ should be replaced by  $P_2^+(z_0) + s^2 \mathbb{P}(Z(t_1)>0, Z(t_2)<0|Z(0)=z_0)$. 
    We note that all even annealed cumulants $\kappa_{\mathrm{even},k}^A$ (and all odd annealed cumulants $\kappa_{\mathrm{odd},k}^A$) involving $k$ times are equal.

    As in the unbiased case, we compute all cumulants up to order $4$. The lowest one $\kappa_{\mathrm{odd}, 1}^{A} \sim \kappa_1^{Q} \sim s\sqrt{2t/\pi}$ being well known~\cite{Illien2013,Poncet2021},  we first focus on the covariance $\kappa_2$ in the annealed and quenched settings:
    \begin{gather}
        \kappa_{\mathrm{even}, 2}^A \sim \frac{1}{\sqrt{2\pi}}\left[u_+\sqrt{t_1} + u_-(\sqrt{t_2} - \sqrt{t_2-t_1})\right], \\
        \kappa_{2}^Q \sim \frac{1}{\sqrt{2\pi}}\left[u_+\sqrt{t_1+t_2} - u_-\sqrt{t_2-t_1} -2s^2\sqrt{t_2}\right],
    \end{gather}
    with $u_\pm = 1\pm s^2$. These expressions are plotted in Fig.~\ref{fig:biased}b and call for several comments.
    (i) As soon as $s\neq 0$, the annealed process no longer corresponds to a fractional Brownian motion.
    (ii) In the annealed case, the memory effect at large time (right side of Fig.~\ref{fig:biased}b) changes with the magnitude of bias. Strikingly, at $s=\pm1$, the increments become uncorrelated, $\langle X(t_1)[X(t_2)-X(t_1)]\rangle = 0$, leading to a covariance that no longer depends on $t_2$. (iii) In contrast with the annealed case, the  quenched covariance always decays to $0$ at large $t_2$ (fixed $t_1$). Note that this decay is almost independent of the bias as soon as $t_2>2t_1$: the main variation with the bias is encoded in the value when $t_2=t_1$: 
    the single-time variance $\kappa_{(11)}^Q$~\cite{Poncet2021}.

    While this phenomenology of a biased tracer is already rich at the Gaussian order, our framework allows us to   go beyond  and investigate the 3-time correlations, leading to
    \begin{multline}
        \kappa_{\mathrm{odd},3}^A \sim \frac{s}{2\sqrt{2\pi}}\bigg\{4\sqrt{t_1} 
        +u_- \bigg[(a_{123}-1)\sqrt{t_1} + \sqrt{t_2} \\ + (b_{0123}-1)\sqrt{t_3} 
        - \sqrt{t_2-t_1} + \sqrt{t_3-t_1} - a_{012} \sqrt{t_3-t_2}\bigg]\bigg\}.
    \end{multline}
    When $t_2=t_3$, this expression reduces to $\kappa_{\mathrm{odd},2}^A(t_1, t_2) \sim s\sqrt{2t_1/\pi}$ which surprisingly is independent of $t_2$.
    In End matter, we provide the full expression for the 4-time annealed cumulant  $\kappa_{\mathrm{even},4}$. Its structure is similar to Eq.~\eqref{eq:kappaAeven4}, with coefficients depending on $s^2$.
    The quenched cumulants can also be computed:
    Fig.~\ref{fig:biased}c shows the evolution of 
    $\kappa^Q_{(122)}$ with $t_2-t_1$ and $s$. Contrary to the covariance, the sign varies and the behavior is non-monotonous.

    \textit{Beyond the large time limit.} Last, we highlight that our approach also enables us to investigate the problem at all times, and go beyond the large-time limit. These results are best written in terms of the correlations of increments $K_n(\vvec\tau)=\lim_{\rho\to 1} \frac{1}{1-\rho} \langle \prod_{i=1}^n [X(t_i) - X(t_{i-1})] \rangle_c$ with the time increments 
    $\vvec\tau=\{\tau_i\}, \tau_i = t_i-t_{i-1}$.
    The Laplace transform, $\hat K_n(\vvec u) = \int_0^\infty d\vvec \tau e^{-\vvec u\cdot\vvec \tau} K_n(\vvec\tau)$, takes a compact form for an unbiased tracer in the annealed setting. The lowest  order, $\hat K_1^A(u_1) = [u_1^3(2+u_1)]^{-1/2}$,  already computed in  ~\cite{Poncet2022}, can be inverted to obtain  analytically the diffusive to subdiffusive crossover. Next, 
    \begin{equation} \label{eq:K2Au}
        \hat K_2^A = \frac{-1}{u_1u_2}\frac{1}{(2+u_1)\sqrt{u_2(2+u_2)} + (2+u_2)\sqrt{u_1(2+u_1)}},
    \end{equation}
     which gives back the  limits of Brownian motion at short times ($K_2(\vvec\tau) = o(\vvec \tau)$) and fractional Brownian motion at large times (Eq.~\eqref{eq:kappa2Ares}), provides after numerical Laplace inversion the behavior at all times.
    The Laplace transform $\hat{K}_n^A$ is provided in End Matter. As emphasized previously, the expression of $K_n$ actually holds for cumulants of arbitrary orders involving $n$ times.

    \textit{Conclusion.---} We studied the full stochastic process of a tracer in the dense SEP beyond the Gaussian order, deriving both the structure of the correlations and their explicit expressions up to order $4$. We tackled important extensions such as a step initial condition and a bias on the tracer. Our work constitutes a rare instance of an explicit non-Markovian and non-Gaussian stochastic process. We highlighted the strong memory effects induced by both the initial conditions and the bias, and traced them to conditional probabilities of a Brownian walker.
    By shedding light on these fundamental aspects, our study provides, along with Ref.~\cite{Sadhu_2015}, a step toward a more complete understanding of non-Markovian processes in exclusion models at arbitrary density.
    Beyond that, we stress that our high-density framework 
    is generic and should be applicable to study stochastic processes in other geometries, such as comblike structures~\cite{Benichou2015,Iomin2018,Venturelli2025} and higher-dimensional lattices~\cite{Spohn1991,Venturelli2025b}.


    \bibliographystyle{apsrev4-2}

    \onecolumngrid

    \vspace{12pt}
    \noindent\hrulefill \hspace{24pt} {\bf End Matter} \hspace{24pt} \hrulefill
    \vspace{12pt}
    
    \begin{table*}[h!]
        \centering
        \begin{tabular}{c|c||c|c||c|c}
		$C^{(4,e)}_{01}$ & $8s^2 + u_-\left(u_- + u_+a_{123}+u_-a_{124}+u_+a_{134}\right)$ & 
		$C^{(4,e)}_{02}$ & $u_-\left(u_+ + u_-a_{234}\right)$ &
		$C^{(4,e)}_{03}$ & $u_-\left(u_-+u_+b_{0123}\right)$  \\
		$C^{(4,e)}_{04}$ & $u_-\left(u_+ + u_-b_{0124} + u_+ b_{0134} + u_- b_{0234}\right)$ &
		$C^{(4,e)}_{12}$ & $-u_-\left(u_+ + u_-a_{234}\right)$ &
		$C^{(4,e)}_{13}$ & $-u_-^2$ \\
		$C^{(4,e)}_{14}$ &  $-u_-\left(u_+ + u_-b_{1234}\right)$ &
		$C^{(4,e)}_{23}$ &  $-u_-u_+ a_{012}$ &
		$C^{(4,e)}_{24}$ &  $-u_-^2 a_{012}$ \\
		$C^{(4,e)}_{34}$ &  $-u_-\left(u_+ a_{013} + u_- a_{023} - u_- a_{123}\right)$ &
		$\tilde C^{(4,o)}_{01}$ & $4u_+ +2u_-\left(a_{123}+a_{134}\right)$ & 
		$\tilde C^{(4,o)}_{02}$ & $2u_-$ \\
		$\tilde C^{(4,o)}_{03}$ & $2u_-b_{0123}$  &
		$\tilde C^{(4,o)}_{04}$ & $2u_-\left(b_{0134} - 1\right)$ &
		$\tilde C^{(4,o)}_{12}$ & $-2u_-$ \\
		$\tilde C^{(4,o)}_{14}$ &  $2u_-$ &
		$\tilde C^{(4,o)}_{23}$ &   $-2u_- a_{012}$ &
		$\tilde C^{(4,o)}_{34}$ &  $-2u_- a_{013}$
        \end{tabular}
        \caption{Annealed cumulants of a biased tracer. We write $C^{(4,o)}_{ij} =s\tilde C^{(4,o)}_{ij}$ and note that $\tilde C^{(4,o)}_{13}=\tilde C^{(4,o)}_{24}=0$. The expressions of $a_{ijk}$ and $b_{ijkl}$ are provided in the main text, and  $u_\pm = 1\pm s^2$.}
        \label{tab:annealed}
    \end{table*}

    \begin{table*}[h!]
        \centering
        \resizebox{\textwidth}{!}{
        \begin{tabular}{c|c||c|c||c|c||c|c}
		$\tilde D^{(4)}_{12}$ & $-a_{234}$ &
		$\tilde D^{(4)}_{13}$ &  $0$ &
		$\tilde D^{(4)}_{14}$ &  $-b_{1234}$ &
		$\tilde D^{(4)}_{23}$ &  $1-a'_{421}$ \\
		$\tilde D^{(4)}_{24}$ &  $1-a'_{312}$  &
		$\tilde D^{(4)}_{34}$ &  $1+a_{123}-a'_{123}-a'_{213}$ &
		$\tilde E^{(4)}_{12}$ &  $a_{134}+a_{234}$ &
		$\tilde E^{(4)}_{13}$ &  $a_{124}-2g_{1324}$ \\
		$\tilde E^{(4)}_{14}$ &  $a_{123}+e'_{1423}$ &
		$\tilde E^{(4)}_{23}$ &  $-2g_{2314}-2g_{3214}$ &
		$\tilde E^{(4)}_{24}$ &  $e''_{2413}-2g_{4213}$ &
		$\tilde E^{(4)}_{34}$ &  $e''_{3412}+e''_{4312}$ \\ 
        \hline
        $\tilde D_{01}^{(3,\sigma)}$ & $a_{123}-a_{023}$ &
        $\tilde D_{02}^{(3,\sigma)}$ & $-a_{013}$ &
        $\tilde D_{01}^{(3,s)}$ & $u_-\left[a_{123}-a_{012}\right]$ &
        $\tilde D_{02}^{(3,s)}$ & $-u_-a_{013}$ \\
        $\tilde D_{03}^{(3,\sigma)}$ & $b_{0123}-a_{012}$ & 
        $\tilde D_{12}^{(3,\sigma)}$ & $-p_{13}$ &
        $\tilde D_{03}^{(3,s)}$ & $4s^2 + u_-\left[b_{0123} - a_{012} \right]$ &
        $\tilde D_{12}^{(3,s)}$ & $-u_- p_{13}$ \\
        $\tilde D_{13}^{(3,\sigma)}$ & $-p_{12}$ &
        $\tilde D_{23}^{(3,\sigma)}$ & $p_{12}-a_{012}$ &  
        $\tilde D_{13}^{(3,s)}$ & $u_-\left[2 - p_{12}\right]$ &
        $\tilde D_{23}^{(3,s)}$ & $u_-\left[p_{12}-a_{012}\right]$ \\ 
        $\tilde E_{12}^{(3,\sigma)}$ & $2n_{123}$ &
        $\tilde E_{13}^{(3,\sigma)}$ & $2n_{132}-m_{123}$ &
        $\tilde E_{12}^{(3,s)}$ &  $2v\left[1-n_{312}\right]$ &
        $\tilde E_{13}^{(3,s)}$ & $4 - u_- m_{123} -2v n_{213}$ \\
        $\tilde E_{23}^{(3,\sigma)}$ & $2n_{231}-m_{213}-m_{312}$ & & &
        $\tilde E_{23}^{(3,s)}$ & $u_-\left[2-m_{213}-m_{312}\right]-2v n_{123}$  & & \\
        \hline
        $a_{ijk}$ & $A\left(\frac{t_j-t_i}{t_k-t_j}\right)$ &
        $a'_{ijk}$ & $A\left(\frac{t_i+t_j}{t_k-t_j}\right)$ &
        $b_{ijkl}$ & $A\left(\frac{(t_j-t_i)(t_l-t_k)}{(t_l-t_i)(t_k-t_j)}\right)$ &
        $c_{ijk}$ & $A\left(\frac{t_i^2}{t_it_j+t_it_k+t_jt_k}\right)$ \\
        $d_{ijkl}$ & $A\left(\frac{t_it_j+t_it_k+t_jt_k}{(t_i+t_j)(t_l-t_k)}\right)$ &
        $e_{ijkl}$ & $A\left(\frac{(t_j-t_l)(t_i+t_k)}{(t_l-t_k)(t_i+t_j)}\right)$ &
        $e'_{ijkl}$ & $e_{ijkl}-2g_{ijkl}-2g_{ijlk}$ &
        $e''_{ijkl}$ & $e'_{ijkl}+f_{ijkl}$ \\
        $f_{ijkl}$ & $A\left(\frac{(t_k-t_i)(t_l-t_j)}{(t_i+t_j)(t_k+t_l)}\right)$ &
        $g_{ijkl}$ & $A\left(\frac{t_i^2(t_j-t_k)}{(t_i+t_j)(t_it_k+t_it_l+t_kt_l)}\right)$ &
        $h_{ijkl}$ & $A\left(\frac{t_i^2t_j^2/(t_i+t_j)}{t_it_j(t_k+t_l) + t_kt_l(t_i+t_j)}\right)$ & &
        \\
        $p_{ij}$ & $A\left(\frac{t_i}{t_j}\right)$ &
        $n_{ijk}$ & $A\left(\frac{t_it_j}{t_k(t_i+t_j)}\right)$ &
        $m_{ijk}$ & $A\left(\frac{t_i(t_k-t_j)}{t_j(t_i+t_k)}\right)$ 
        \end{tabular}
        }
        \caption{Quenched cumulants. 
        First part: equilibrium SEP. We have $D^{(4)}_{0j} = 0$ and $D^{(4)}_{ij} = a'_{ikl}-2c_{ikl} + \tilde D^{(4)}_{ij}, E^{(4)}_{ij}=-2d_{ijkl}+6h_{ijkl} + \tilde E^{(4)}_{ij}$ for $1\leq i < j \leq 4$, with $1\leq k<l\leq 4$ the two indices different from $i$ and $j$. $\tilde D^{(4)}_{ij}$ and $\tilde E^{(4)}_{ij}$ are listed in the second part of the table. 
        Second part: coefficients $D_{ij}^{(3,\sigma)}=\sigma\tilde D_{ij}^{(3,\sigma)}$ for a step initial profile $\sigma$, and $D_{ij}^{(3,s)}=s\tilde D_{01}^{(3,s)}$ for a biased tracer (and similarly for $E_{ij}^{(3)}$).
        Last part: definition of the quantities introduced in terms of the function $A(u) = ({2}/{\pi})\arctan\sqrt{u}$. We additionally used $u_\pm = 1\pm s^2$ and $v=3+s^2$.
        }
        \label{tab:quenched}
    \end{table*}

    \twocolumngrid

    {\it Appendix A: Explicit expressions of cumulants.---} The annealed cumulants of even order have the general expression,
    \begin{equation}
        \kappa_{\mathrm{even}, k}^A \sim \frac{2^{2-k}}{\sqrt{2\pi}}\sum_{0\leq i < j \leq k} C^{(k, e)}_{ij} \sqrt{t_j-t_i},
    \end{equation}
    and similarly for the odd cumulants $\kappa_{\mathrm{odd}, k}^A$ with coefficients $C^{(k, e)}_{ij}$. The coefficients for $k=4$ when the tracer is biased are provided in Table~\ref{tab:annealed}.

    The quenched cumulants are as follow.
    \begin{equation}
    \kappa_n^Q \sim \frac{2^{2-k}}{\sqrt{2\pi}}\left[\sum_{0\leq i<j\leq n} \!\!\!\! D_{ij}^{(n)} \sqrt{t_j-t_i} +\!\!\! \sum_{1\leq i<j\leq n}\!\! E_{ij}^{(n)} \sqrt{t_i+t_j}\right]    
    \end{equation}
    In Table~\ref{tab:quenched}, we provide the coefficients in three cases:
    $n=4$ for the equilibrium SEP, $n=3$ for a step initial condition, and $n=3$ for a biased tracer. The Mathematica notebooks for computing the cumulants are provided in Ref.~\cite{Poncet_TracerDenseSingleFile_2025_2}.

    {\it Appendix B: Cumulants at all times.---} The Laplace transform of the correlations of increments takes a compact form for an unbiased tracer in the annealed setting:
    \begin{equation}
        \hat K_{2n}^A(\{u_i\}) = 2\prod_{j=1}^{2n} \frac{\alpha_j}{u_j(1+\alpha_j)} \frac{\prod_{j=2}^{2n-1}(1-\alpha_j)}{\prod_{j=1}^{2n-1}(1-\alpha_j\alpha_{j+1})},
    \end{equation}
    with $\alpha_j = 1+u_j-\sqrt{u_j(2+u_j)}$. For $n=1$, this Laplace transform can be inverted~\cite{Poncet2022} and one obtains
    \begin{equation}
        K_1(\tau) = \tau e^{-\tau}[I_0(\tau) + I_1(\tau)]
    \end{equation}
    where $I_n$ are modified Bessel functions. This provides analytically the crossover from diffusion to subdiffusion.
    
    {\it Appendix C: Edwards-Wilkinson equation (EW).---} 
    We define the EW equation for an interface $h(z, t)$ as
     \begin{equation} \label{smeq:ew}
        \partial_t h(z, t) = \frac{1}{2} \partial_{zz} h(z, t) + \sqrt{D} \eta(z, t)n
    \end{equation}
    with $\eta$ a Gaussian white noise of vanishing average and correlation $\langle \eta(z, t)\eta(z',t')\rangle = \delta(z-z')\delta(t-t')$.
    The general solution of Eq.~\eqref{smeq:ew} starting at $t=0$ from a flat interface $h(z, 0)=0$ is
    \begin{equation}
        h(z, t) = \sqrt{D} \int_0^t dt_1 \int_{-\infty}^\infty dz_1 \eta(z_1, t_1) \frac{e^{-\frac{(z-z_1)^2}{2(t-t_1)}}}{\sqrt{2\pi(t-t_1)}} .
    \end{equation}
    The two-point correlation $c_Q(z, t, \tau) = \langle h(0, t) h(z, t+\tau)\rangle$ reads
    \begin{align} \label{eq:ew_cq}
        c_Q = \frac{D}{\sqrt{2\pi}}\left[ g\left(\frac{|z|}{\sqrt{2(2t+\tau)}}\right)\sqrt{2t+\tau} -   g\left(\frac{|z|}{\sqrt{2\tau}}\right)\sqrt{\tau}\right], 
    \end{align}
    with $g(z) = e^{-z^2} - \sqrt{\pi} z \erfc(z)$. By inputing $D=1-\rho$, we obtain the quenched covariance of the dense SEP.
    
    We now set the initial condition at $t=-T$, with $T\to \infty$. The covariance starting from an equilibrated interface (at time $t=0$) is
    $c_A(z, t, \tau) = \left\langle[h(z, t+\tau) - h(z, 0)][h(0, t) - h(0, 0)]\right\rangle$. 
    Using Eq.~\eqref{eq:ew_cq},
    \begin{multline}
        c_A(z, t, \tau) = \frac{D}{\sqrt{2\pi}}\bigg[ g\left(\frac{|z|}{\sqrt{2t}}\right)\sqrt{t}\\ + g\left(\frac{|z|}{\sqrt{2(t+\tau)}}\right)\sqrt{t+\tau}  -   g\left(\frac{|z|}{\sqrt{2\tau}}\right)\sqrt{\tau}\bigg]. 
    \end{multline}
    If $D=1-\rho$, we recover the annealed covariance of the dense  SEP.

    The limitations of the EW equation in the present context are clear: (i) the noise coefficient $D$ needs to be inputed by hand, (ii) it is a Gaussian model unable to reproduce cumulants of order higher than 2.

\clearpage
\widetext

\let\addcontentsline\oldaddcontentsline

\begin{center}
  \begin{large}

    \textbf{
     Supplemental Material for\texorpdfstring{\\}{}
     Full stochastic dynamics of a tracer in a dense single-file system
   }
  \end{large}
   \bigskip

   Alexis Poncet, Aurélien Grabsch and Olivier B\'enichou
\end{center}

\setcounter{equation}{0}
\renewcommand{\theequation}{S\arabic{equation}}
\renewcommand{\thefigure}{S\arabic{figure}}
\renewcommand{\bibnumfmt}[1]{[S#1]}
\renewcommand{\citenumfont}[1]{S#1}
\setcounter{secnumdepth}{3}

\tableofcontents

    \section{Dense limit of the symmetric exclusion process}
    \subsection{Symmetric exclusion process}
    The symmetric exclusion process is a stochastic model on the infinite discrete line defined as follow. Initially, each site is occupied independently by a particle, with probability $\rho$ where $\rho$ is the density. The particles are embedded with independant clocks governed by exponential laws $\Xi(t)=e^{-t}$. When its clock rings, a particle chooses uniformly at random between its left and right neighboring site. If this site is empty, the particle jumps to it, overwise the move is rejected. While the collective dynamics of the SEP is diffusive, the motion of a given particle is highly non-trivial. It is this stochastic process that we investigate in this article.
    
    \subsection{Cumulant generating functions and n-time cumulants}
    \subsubsection{Definitions}
    Let us consider an tracer in the SEP. It is initially at the origin and we denote $X(t)$ its position at time $t$. Our main goal is to characterize the stochastic process $\{X(t)\}_{t\geq 0}$. This amounts to obtaining the $n$-time moments $\langle X(t_1)\dots X(t_n)\rangle$ for all times $0<t_1\leq t_2\leq \dots\leq t_n$, for all $n\geq 1$. We therefore define the $n$-time cumulant generating function (CGF),
    \begin{equation} \label{smeq:def_chi}
        \tilde \chi_n(\mu_1, \dots, \mu_n, t_1, \dots t_n) = \ln\left\langle e^{\mu_1 X(t_1)+\dots +\mu_n X(t_n)}\right\rangle.
    \end{equation}
    As its name indicates, this function generates the $n$-times connected correlations (denoted $\langle\dots\rangle_c$), also known as cumulants,
    \begin{equation}
        \langle X(t_{i_1})\dots X(t_{i_k})\rangle_c = \left.\frac{\partial^k \tilde \chi_n}{\partial\mu_{i_1}\dots \partial\mu_{i_k}}\right|_{\mu_1=\dots=\mu_n=0}.
    \end{equation}
    If one allows consecutive times to be equal, it is enough to focus of $\tilde\kappa_n = \langle X(t_1)\dots X(t_n)\rangle_c$. The simplest example is the covariance $\tilde\kappa_2(t_1, t_2) = \langle X(t_1)X(t_2)\rangle - \langle X(t_1)\rangle\langle X(t_2)\rangle$. The $4$-time cumulants is $\tilde\kappa_4(t_1, t_2, t_3, t_4) = \langle\bar X_1\bar X_2\bar X_3\bar X_4\rangle - \tilde\kappa_2(t_1, t_2)\tilde\kappa_2(t_3, t_4)-\tilde\kappa_2(t_1, t_3)\tilde\kappa_2(t_2, t_4)-\tilde\kappa_2(t_1, t_4)\tilde\kappa_2(t_2, t_3)$ with $\bar X_i = X(t_i) - \langle X(t_i)\rangle$.

    The framework that we build in the following needs an equivalent description in term of the increments $\Delta X_i = X(t_{i+1})-X(t_i)$ (with the convention $t_0=0$). We  introduce the CGF of increments
    \begin{equation} \label{smeq:def_psi}
        \tilde \psi_n(\lambda_1, \dots, \lambda_n, t_1, \dots t_n) = \ln\left\langle e^{\lambda_1\Delta X_1+\dots +\lambda_n\Delta X_n}\right\rangle.
    \end{equation}
    One easily sees that the two descriptions are related by a simple change of variables: $\chi_n=\psi_n$ with $\mu_i = \lambda_i-\lambda_{i+1}$ ($\lambda_{n+1}=0$).
    The cumulants of the increments increments are $\tilde K_n = \langle \Delta X_1\dots\Delta X_n\rangle_c$.
    
    \subsubsection{Side comment: fractional Brownian motion}
    Since the reader may not be familiar with continuous time stochastic processes, we provide a simple example: the fractional Brownian motion with Hurst index $H$. It is a Gaussian process (characterized only by its averages and covariances) that has zero mean and stationary increments. In details, the $n$-point cumulants are
    \begin{align}
        \langle X(t_1) X(t_2)\rangle_c &= \frac{1}{2}\left(t_1^{2H} + t_2^{2H} - |t_2-t_1|^{2H}\right), \\
        \langle X(t_1)\dots X(t_n)\rangle_c &= 0 \qquad \text{if } n\neq 2.
    \end{align}
    The CGF reads
    \begin{equation}
        \chi(\mu_1, \dots, \mu_n) = \sum_{i, j}\tilde \kappa_{(ij)} \mu_i\mu_j,
    \end{equation}
    with $\tilde \kappa_{(ij)} = \langle X(t_i) X(t_j)\rangle_c$.
    
    The case $H=1/2$ corresponds to the usual Brownian motion, while $H=1/4$ has been show to be the Gaussian limit (no correlations for more than two times) of the motion of an tracer in the SEP~\cite{peligrad_2008_SM}. Our main goal is to characterize the process beyond this Gaussian limit, and obtain higher-order correlation functions.

    \subsubsection{Annealed and quenched averages}
    In Eqs.~\eqref{smeq:def_chi} and \eqref{smeq:def_psi}, we did not explicit what the averaging means. There are two part to it: averaging over the initial conditions of the system, that is to say the initial positions of the particles in the SEP, and averaging over the stochastic evolution of the system starting from  a given initial condition. In what follows, we shall distinguish between the annealed and quenched averages. The annealed CGF,
    \begin{equation}\label{smeq:def_psiA}
        \tilde \psi_n^A(\lambda_1, \dots, \lambda_n, t_1, \dots t_n) = \ln\left \langle\left\langle e^{\lambda_1\Delta X_1+\dots +\lambda_n\Delta X_n}\right\rangle_\mathrm{ev}\right\rangle_\mathrm{ic},
    \end{equation}
     corresponds to an average at the same time over both the stochastic evolution ($\mathrm{ev}$) and the initial conditions ($\mathrm{ic}$). On the other hand, the quenched CGF
    \begin{equation}\label{smeq:def_psiQ}
        \tilde \psi_n^Q(\lambda_1, \dots, \lambda_n, t_1, \dots t_n) = \left \langle\ln\left\langle e^{\lambda_1\Delta X_1+\dots +\lambda_n\Delta X_n}\right\rangle_\mathrm{ev}\right\rangle_\mathrm{ic}.
    \end{equation}
    corresponds to computing the CGF for a given initial condition and then averaging  the CGFs for all initial conditions. This corresponds to the ``typical'' CGF for fixed initial conditions.

    The SEP exhibits a strong memory of the initial conditions that is reflected in the quantitative and qualitative differences between $\tilde \psi_n^A$ and $\tilde \psi_n^Q$ shown in the main text.

    \subsection{High-density CGF from the single-vacancy problem}
    Let us now consider a dense system of size $N$ with $M$ empty sites ($N-M$ sites    are occupied by a particle) with $M/N = \rho_0\to 0$, where $\rho_0=1-\rho$ and $\rho$ is the density of the system. Rather than looking at the particles, we focus on these empty sites that we call vacancies. The displacement $Y_k$ of the tagged particle between times $t_{k-1}$ and $t_k$ is written as $Y_k=Y_k^1+\dots+Y_k^M$ where $Y_k^j$ is the displacement due to the $j$-th vacancy (that is to say the combination of the events where the vacancy and the particle exchanged their positions). This statement readily translates into the relation
    \begin{equation} \label{smeq:PnY}
        \mathcal{P}^{(n)}(Y_1, \dots, Y_n, \tau_1, \dots, \tau_n|\{Z_j\}) = \sum_{Y^1_1, \dots, Y^M_1}\delta_{Y_1, Y^1_1+\dots Y^M_1}  \dots \sum_{Y^1_n, \dots, Y^M_n}\delta_{Y_n, Y^1_n+\dots Y^M_n}   \mathcal{P}^{(n)}_\mathrm{vac}(\{Y_1^j\}, \tau_1, \dots \{Y_n^j\}, \tau_n|\{Z_j\})
    \end{equation}
    where $\mathcal{P}^{(n)}(Y_1, \dots, Y_n, \tau_1, \dots, \tau_n|\{Z_j\})$ is the probability that the TP have a displacement $Y_k$ during time increment $\tau_k=t_k-t_{k-1}$ knowing that the $M$ vacancies where initially at positions $Z_1, \dots, Z_M$; and $\mathcal{P}^{(n)}_\mathrm{vac}(\{Y_1^j\}, \tau_1, \dots \{Y_n^j\}, \tau_n|\{Z_j\})$ is the probability that vacancy $j$ induced a displacement $Y^j_k$ of the TP (for all $j$) in the same setup.

    At high density ($M/N\to 0$), the vacancies become independent of one another, and neglecting events of order $\mathcal{O}[(1-\rho)^2]$, the probabilities defined above factorize \cite{Illien2013_SM,Poncet2021_SM},
    \begin{align} \label{smeq:Pvac}
        \mathcal{P}^{(n)}_\mathrm{vac}(\{Y_1^j\}, \tau_1, \dots \{Y_n^j\}, \tau_n|\{Z_j\}) &\equi{\rho\to 1} \prod_{j=1}^M P^{(n)}(Y^j_1, \dots,  Y^j_n, \tau_1, \dots, \tau_n| Z_j).
    \end{align}
    $P^{(n)}(Y_1, \dots, Y_n, \tau_1, \dots, \tau_n| Z)$ corresponds to the problem where there is a \emph{single} vacancy in the system ($M=1$). It is the probability that the tracer has  successive displacements $Y_1$, \dots, $Y_n$ during time increments $\tau_1, \dots \tau_n$ knowing that the vacancy was initially at position $Z$.

    Using Eqs.~\eqref{smeq:PnY} and \eqref{smeq:Pvac}, and defining the generating function $\tilde f(k)$ of a spatially varying function $f(Y)$ as $\tilde f(\lambda) = \sum_{Y=-\infty}^\infty e^{\lambda Y} f(Y)$, we obtain
    \begin{align} \label{smeq:PnHigh}
        \mathcal{\tilde P}^{(n)}(\lambda_1, \dots, \lambda_n, \tau_1, \dots, \tau_n|\{Z_j\}) &\equi{\rho\to 1} \prod_{j=1}^M \tilde  P^{(n)}(\lambda_1, \dots, \lambda_n, \tau_1, \dots, \tau_n|Z_j). 
    \end{align}
    The generating function $\mathcal{\tilde P}^{(n)}$ corresponds to the average of $e^{\prod_i \lambda_i Y_i}$ over the evolution of the system for fixed initial conditions encoded in $\{Z_j\}_{j=1}^M$. To compute the CGFs , we also need to average over these initial conditions. This averaging, denoted $\langle\bullet \rangle_\mathrm{ic}$, is performed by summing over the $(N-1)$ possible initial positions of each of the $M$ (independent) vacancies: 
    \begin{equation}
        \langle\bullet\rangle_\mathrm{ic} = \frac{1}{(N-1)^M} \sum_{Z_1, \dots, Z_M\neq 0}  \bullet .
    \end{equation}
    In details, the annealed and quenched CGF, Eqs.~\eqref{smeq:def_psiA}-\eqref{smeq:def_psiQ}, are given by
    \begin{align}
        \tilde \psi_n^A(\lambda_1, \dots, \lambda_n, t_1, \dots t_n) &=\log \left\langle\mathcal{\tilde P}^{(n)}(\lambda_1, \dots, \lambda_n, \tau_1, \dots, \tau_n|\{Z_j\})\right\rangle_\mathrm{ic},  \\
        \tilde \psi_n^Q(\lambda_1, \dots, \lambda_n, t_1, \dots t_n) &= \left\langle\log\mathcal{\tilde P}^{(n)}(\lambda_1, \dots, \lambda_n, \tau_1, \dots, \tau_n|\{Z_j\})\right\rangle_\mathrm{ic}.
    \end{align}
    We consider the thermodynamic limit $N\to \infty$ at fixed fraction of vacancies $M/N=1-\rho$. Using Eq.~\eqref{smeq:PnHigh}, the CGFs at high density become proportional to $(1-\rho)$ and we obtain 
	\begin{align}
	   \psi^{A}_n(\lambda_1, \dots, \lambda_n, \tau_1, \dots \tau_n) = \lim_{\rho\to 1} \frac{\tilde \psi^{A}_n(\lambda_1, \dots, \lambda_n, \tau_1, \dots \tau_n)}{1-\rho} &= \sum_{Z\neq 0} \left[ \tilde P^{(n)}(\lambda_1, \dots, \lambda_n, \tau_1, \dots \tau_n)|Z) - 1 \right] \label{eq:psina_hd}, \\
       \psi^{Q}_n(\lambda_1, \dots, \lambda_n, \tau_1, \dots \tau_n) =\lim_{\rho\to 1} \frac{\tilde \psi^{Q}_n(\lambda_1, \dots, \lambda_n, \tau_1, \dots \tau_n)}{1-\rho} &= \sum_{Z\neq 0} \log \tilde P^{(n)}(\lambda_1, \dots, \lambda_n, \tau_1, \dots \tau_n)|Z). \label{eq:psinq_hd}
	\end{align}

    \subsection{Propagator in the single-vacancy system} \label{ss:prop}
    We first remark that the propagator $P^{(n)}$ is the single vacancy system can be decomposed on the successive times $t_1, \dots, t_n$. We denote $Z_1, \dots, Z_n$ the position of the vacancy at these times. The decomposition reads:
    \begin{align}
			P^{(n)}(Y_1, \dots, Y_n, \tau_1, \dots \tau_n|Z_0) &= \sum_{Z_{n}, \dots Z_1\neq 0}  P^\dagger(Y_n, \tau_n, Z_n|Z_{n-1}) \dots P^\dagger(Y_1, \tau_1, Z_1|Z_0), \label{smeq:PnDecomp} \\
			\tilde P^{(n)}(\lambda_1, \dots, \lambda_n, \tau_1, \dots \tau_n|Z_0) &= \sum_{Z_{n}, \dots Z_1\neq 0} \tilde P^\dagger(\lambda_n, \tau_n, Z_n|Z_{n-1}) \dots \tilde P^\dagger(\lambda_1, \tau_1, Z_1|Z_0) . \label{smeq:PnDecompF} 
    \end{align}
    The problem reduces to the computation of $P^\dagger(Y, \tau, Z_1|Z_0)$ which is the probability that the tracer has a displacement $Y$ during time $\tau$, knowing the positions of the vacancy at both initial time, $Z_0$, and final time, $Z_1$. $\tilde P^\dagger(\lambda, \tau, Z_1|Z_0)$ is the generating function of this probability.

    In the single-vacancy system, the TP only moves when it is touched by the vacancy, that is to say when the vacancy comes to the origin. As a consequence, we only need to study the random walk performed by the vacancy. Let us assume in this section that the following two quantities are known. Their computation in specific cases is delayed to the next sections.
   \begin{enumerate}
       \item The probability $f_{Z_1, Z_0}(\tau)$ of \emph{first passage} at position $Z_1$ and at time $\tau$ of the vacancy that was at initial position $Z_0$ at time $0$. As we are interested in returns of the vacancy at the origin, we will use the short-hand notation $f_Z(\tau) = f_{0, Z}(\tau)$.
       \item The probability $\Pi_{Z_1,Z_0}(\tau)$ that the vacancy moves from $Z_0$ to $Z_1$ during a time interval $\tau$ \emph{without touching the origin}.
   \end{enumerate}
   
   Once we know these two quantities, the propagator $P^\dagger(Y, \tau, Z_1|Z_0)$ can be computed by partitioning on whether the vacancy touches the origin or not:
    \begin{align}
        P^\dagger(Y, \tau, Z_1|Z_0) &= \delta_{Y, 0} \Pi_{Z_1, Z_0}(\tau) + \int_0^\tau d\tau_0 f_{0, Z_0}(\tau_0) P^\dagger(Y-\nu_0, \tau-\tau_0, Z_1|-\nu_0), \\
        \hat{\tilde P}^\dagger(\lambda, u, Z_1 | Z_0) &= \hat\Pi_{Z_1, Z_0}(u) + e^{\nu_0 \lambda}
	\hat{\tilde P}^\dagger(\lambda, u, Z_1|-\nu_0) \hat f_{Z_0}(u).
    \label{eq:Pdagger1}
    \end{align}
    The second line is the generating function in space and Laplace transform in time of the first one, with the definition $\hat g(u) = \int_0^\infty d\tau e^{-u\tau} g(\tau)$. We denote $\nu_0=\pm 1$ the sign of $Z_0$.

    The goal is thus to compute $P^\dagger(Y, \tau, Z_1|\mu)$ for $\mu=\pm 1$. It can be done by summing over the $n\geq 0$ successive returns of the vacancy at the position of the TP:
    \begin{align}
        P^\dagger(Y, \tau, Z_1|\mu) = \sum_{n=0}^\infty \delta_{Y, \mu[1-(-1)^{n+1}]} \int_{0}^\infty d\tau_1\dots d\tau_{n+1} \delta\left(\tau-\sum_{k=1}^{n+1} \tau_k \right) f_\mu(\tau_1) f_{-\mu}(\tau_2)\dots f_{\mu(-1)^{n-1}}(\tau_n) \Pi_{Z_1, \mu(-1)^n}(\tau_{n+1}).
    \end{align}
    The generative function in space and Laplace transform in time is most easily written by distinguishing the cases $\nu_1 = \mu$ and $\nu_1 = -\mu$ where $\nu_1$ is the sign of $Z_1$,
    \begin{equation}
	\hat{\tilde P}^\dagger(\lambda, u, Z_1 | \mu) = \frac{1}{1-\hat f_1(u)\hat f_{-1}(u)}
	\begin{cases}
		\hat\Pi_{Z_1, \mu}(u) & \mbox{if } \nu_1 = \mu \\
		\hat\Pi_{Z_1, -\mu}(u) \hat f_\mu(u) e^{i\mu k}  & \mbox{if } \nu_1 = -\mu
	\end{cases}.
    \end{equation}
    The expression for arbitrary $Z_0$, from Eq.~\eqref{eq:Pdagger1}, is
    \begin{equation}
    \label{eq:Pdagger2}
    \hat{\tilde P}^\dagger(\lambda, u, Z_1|Z_0) = 
		\begin{cases}
			\hat\Pi_{Z_1,Z_0}(u) + \frac{\hat \Pi_{Z_1, \nu_1}(u) \hat f_{-\nu_0}(u)}{1-\hat f_{1}(u) \hat f_{-1}(u)} f_{Z_0}(u)  & \text{ if } \nu_0 = \nu_1 \\
			e^{\nu_0 \lambda} \frac{\hat \Pi_{Z_1, \nu_1}(u)}{1-\hat f_{1}(u) \hat f_{-1}(u)} f_{Z_0}(u) & \text{ if } \nu_0 = - \nu_1
		\end{cases}.        
    \end{equation}
    At the end of the day, the CGFs are given by Eqs.~\eqref{eq:psina_hd}-\eqref{eq:psinq_hd} with $\tilde P^{(n)}$ decomposed in Eq.~\eqref{smeq:PnDecompF} and $\tilde P^\dagger$ is given by Eq.~\eqref{eq:Pdagger2}.



    \section{Unbiased tracer} \label{s:unbiased}
    \subsection{Expression of the propagators}
    In order to express the propagator $\tilde P^\dagger(\lambda, \tau, Z_1|Z_0)$, we need to compute the quantities $f_Z(t)$ and $\Pi_{Z_1, Z_0}(t)$ associated with the random walk of the vacancy.

    The first passage propagator $f_Z(t)$ is a standard quantity~\cite{Hughes1995_SM}. We rederive it in a few lines. We remind the reader that the exponential clock governing the jump of the walker has a law $\Xi(t) = e^{-t}$, or in Laplace domain $\hat\Xi(u) = (1+u)^{-1}$. The first passage propagator over one site is easily obtained by conditioning over the first jump and using the invariance by translation of the walk.
    \begin{equation}
        f_1(t) = \frac{1}{2} \Xi(t) + \frac{1}{2}\int_0^\infty dt_1 dt_2 dt_3 \delta(t-t_1-t_2-t_3) \Xi(t_1) f_1(t_2) f_1(t_3)
    \end{equation}
    The Laplace transform then reads $\hat f_1(u) = \alpha$ with $\alpha=1+u-\sqrt{u(2+u)}$. For $Z>0$, the walker has to come to site~$1$ before touching the origin: 
    \begin{equation}
    f_Z(t) = \int_0^t dt_1 f_1(t_1) f_{Z-1}(t-t_1).
    \end{equation}
    The recursion relation directly gives $\hat f_Z(u) = \alpha^{|Z|}$ (valid for $Z$ of arbitrary sign by symmetry).

    Another standard quantity~\cite{Hughes1995_SM} that we need to compute is the probability $p_Z(t)$ that the walker is the origin at time $t$ (not necessarily for the first time) knowing that it started from $Z$. $p_0(t)$ can be decomposed as
    \begin{equation}
        p_0(t) = 1 - \int_0^t dt_1 \Xi(t_1) + \int_0^\infty dt_1 dt_2 dt_3 \delta(t-t_1-t_2-t_3) \Xi(t_1) f_1(t_2) p_0(t_3).
    \end{equation}
    This gives the Laplace transform $\hat p_0(u) = [u(2+u)]^{-1/2}$. For arbitrary $Z$, one decomposes
    \begin{equation}
        p_Z(t) = \int_0^t dt_1 f_Z(t_1) p_0(t-t_1)
    \end{equation}
    and obtains
    \begin{equation}
        \hat p_Z(u) = \frac{\alpha^{|Z|}}{\sqrt{u(2+u)}}
        = \frac{1}{u}\frac{1-\alpha}{1+\alpha}\alpha^{|Z|}.
    \end{equation}

    Finally, the probability that the walker goes from $Z_0$ to $Z_1$ without touching the origin, $\Pi_{Z_1, Z_0}(t)$ is obtained by partitioning the walk on whether the origin is touched or not:
    \begin{equation}
            p_{Z_1-Z_0}(t) = \Pi_{Z_1, Z_0}(t) + \int_0^t dt_0 p_{Z_1}(t-t_0) f_{Z_0}(t_0) .
    \end{equation}
        In the Laplace domain,
        \begin{equation}
            \hat \Pi_{Z_1, Z_0}(u) = \hat p_{Z_1-Z_0}(u) - \hat p_{Z_1}(u) \hat f_{Z_0}(u).
        \end{equation}
        Injecting the expressions above, we obtain
        \begin{equation}
        \label{eq:piMW}
            \hat \Pi_{Z_1, Z_0}(u) = \frac{1}{u} \frac{1-\alpha}{1+\alpha}\left(\alpha^{|Z_1-Z_0|}-\alpha^{|Z_1| + |Z_0|}\right).
        \end{equation}
        As expected, $\Pi_{Z_1, Z_0}(t) = 0$ if $Z_0$ and $Z_1$ have opposite signs.

    We can now achieve our goal of giving the explicit expression of $\hat{\tilde P}^\dagger$ from Eq.~\eqref{eq:Pdagger2}:
    \begin{equation}
		\hat{\tilde P}^\dagger(\lambda, u, Z_1|Z_0) = \frac{1}{u} \frac{1-\alpha}{1+\alpha} \times
		\begin{cases}
			\alpha^{|Z_0-Z_1|} & \text{ if } \nu_0 = \nu_1 \\
			\alpha^{|Z_0|+|Z_1|-1} e^{\nu_0 \lambda} & \text{ if } \nu_0 = - \nu_1
		\end{cases}. \label{eq:exprPd}
    \end{equation}
    We recall that $\nu_i$ is the sign of $Z_i$ and $\alpha = 1+u - \sqrt{(1+u)^2 -1}$.
    Note that summing over $Z_1$, we recover the propagator $\hat{\tilde P}(\lambda, u|Z_0)$ already obtained in Ref.~\cite{Poncet2022_SM},
    \begin{equation}
		\hat{\tilde P}(\lambda, u|Z_0) = \sum_{Z_1\neq 0} \hat{\tilde P}^\dagger(\lambda, u, Z_1|Z_0) = \frac{1}{u}
		\left\{1 + \frac{e^{\nu_0 \lambda} - 1}{1+\alpha} \alpha^{|Z_0|} \right\}. \label{eq:exprP}
    \end{equation}

    \subsection{Annealed cumulants for increments at arbitrary times}
    Our goal is now to compute the annealed cumulants of increments,
    \begin{equation}
        K_n^A(\tau_1, \dots, \tau_n) = \lim_{\rho\to 1} \frac{\langle\Delta X_1(\tau_1)\dots X_n(\tau_n)\rangle_c^A}{1-\rho}
        = \left.\frac{\partial^n\psi^{A}_n}{\partial\lambda_1\dots\partial\lambda_n}\right|_{\lambda_1=\dots=\lambda_n=0},
    \end{equation}
    with $\Delta X_i = X(t_i)-X(t_{i-1})$ and $\tau_i = t_i-t_{i-1}$.
    From Eqs.~\eqref{eq:psina_hd} and \eqref{smeq:PnDecompF},
    \begin{equation}
     K_n^A(\tau_1, \dots, \tau_n) = \sum_{Z_{0}, \dots Z_n\neq 0} \frac{\partial}{\partial \lambda_1}\left.\tilde P^\dagger(\lambda_1, \tau_1, Z_1|Z_0)\right|_{\lambda_1=0} \dots \frac{\partial}{\partial \lambda_n}\left.  \tilde P^\dagger(\lambda_n, \tau_n, Z_n|Z_{n-1})\right|_{\lambda_n=0}.
    \end{equation}
    Taking the derivative of Eq.~\eqref{eq:exprPd}, we have
    \begin{equation}
        \frac{\partial}{\partial \lambda}\left.\hat{\tilde P}^\dagger(\lambda, u, Z_1|Z_0)\right|_{\lambda=0} = \nu_0\delta(\nu_0+\nu_1) \frac{1}{u}\frac{1-\alpha}{1+\alpha}\alpha^{|Z_0|+|Z_1|-1}.
    \end{equation}
    Odd cumulants $K_{2n+1}^A$ vanish as expected. The Laplace transform of even cumulants read
    \begin{align}
        \hat K_{2n}^A(u_1, \dots, u_n) &= 2\prod_{i=1}^{2n} \frac{1-\alpha_i}{u_i(1+\alpha_i)} \sum_{Z_{0}', \dots Z_{2n}'=1}^\infty \alpha_1^{Z_0'+Z_1'-1} \dots \alpha_{2n}^{Z_{2n-1}'+Z_{2n}'-1}  \\
        &= 2\prod_{j=1}^{2n} \frac{\alpha_j}{u_j(1+\alpha_j)} \frac{\prod_{j=2}^{2n-1}(1-\alpha_j)}{\prod_{j=1}^{2n-1}(1-\alpha_j\alpha_{j+1})}.
    \end{align}
    This is the result that we report in the main text.
    This Laplace transform can be inverted only in the case of a single time, see Ref.~\cite{Poncet2022_SM}: we now focus on the case of large time increments.


    \subsection{Cumulant-generating functions at large times}
    The expression of $\tilde P^\dagger$, Eq.~\eqref{eq:exprPd}, becomes simpler in the large-time limit $t\to\infty$ with $Z_i/\sqrt{t}$ kept constant. In the Laplace domain, the limit is $u\to 0$ with $Z_i\sqrt{u} = \text{const}$. We obtain
    \begin{align}
        \hat{\tilde P}^\dagger(\lambda, u, Z_1| Z_0) &\equi{u\to 0} \frac{1}{\sqrt{2u}} e^{-|Z_0-Z_1|\sqrt{2u}} e^{\frac{\nu_0-\nu_1}{2}\lambda}, \\
        \tilde P^\dagger(\lambda, t, Z_1| Z_0) &\equi{t\to \infty} \frac{1}{\sqrt{2\pi t}} e^{-\frac{(Z_0-Z_1)^2}{2t}} e^{\frac{\nu_0-\nu_1}{2}\lambda}.
    \end{align}
    Using the decomposition, Eq.~\eqref{smeq:PnDecompF},
    \begin{align}
        \tilde P_n(\{\lambda_i, \tau_i\}|Z_0) &\equi{\tau_i\to\infty} \sum_{Z_1, \dots, Z_n \neq 0} \prod_{i=1}^n \frac{1}{\sqrt{2\pi \tau_i}} e^{-\frac{(Z_{i-1}-Z_i)^2}{2\tau_i}} e^{\frac{\nu_{i-1}-\nu_i}{2}\lambda_i}
    \end{align}
    We consider a large timescale $T\to\infty$ (for instance $T=t_1$), and write $Z_i = z_i\sqrt{2T}$. The sums become integrals:
    \begin{align}
    \label{smeq:PtnScaled}
        \tilde P_n(\{\lambda_i, \tau_i\}|z_0=Z_0/\sqrt{2T}) &\equi{T\to\infty} 
        \int_{-\infty}^\infty \prod_{i=1}^n dz_i K_{\tau_i/T}(z_i, z_{i-1}) e^{\frac{\nu_{i-1}-\nu_i}{2}\lambda_i},
    \end{align}
    with the Gaussian kernel
    \begin{equation}
        K_\tau(z', z) = \frac{1}{\sqrt{\pi\tau}} e^{-\frac{(z'-z)^2}{\tau}}.
    \end{equation}
    The annealed and quenched CGF, Eqs.~\eqref{eq:psina_hd}-\eqref{eq:psinq_hd}, can now be expressed as
    \begin{align} \label{smeq:psiAscaled}
        \psi_n^A(\{\lambda_i, \tau_i\}) &\equi{T\to\infty} \sqrt{2T} \int_{-\infty}^\infty dz_0\left[\tilde P_n(\{\lambda_i, \tau_i\}|z_0) - 1\right] \\
        \label{smeq:psiQscaled}
        \psi_n^Q(\{\lambda_i, \tau_i\})&\equi{T\to\infty} \sqrt{2T} \int_{-\infty}^\infty dz_0\log \tilde P_n(\{\lambda_i, \tau_i\}|z_0).
    \end{align}
    In the following, we formally set $T=1$ (the scaling limit is $\tau_i\to\infty$), this corresponds to the expressions given in the main text. 


    \subsection{Cumulants at large times from a Brownian motion}
    Let us now focus on the large-time limit of the cumulants of the tracer's position
    \begin{equation}\label{smeq:defKappa}
        \kappa_n^{A/Q}(t_1, \dots, t_n) = \lim_{\rho\to 1} \frac{\langle X(t_1)\dots X(t_n)\rangle_c^{A/Q}}{1-\rho} = \left.\frac{\partial^n \chi_n^{A/Q}}{\partial\mu_1\dots\partial\mu_n}\right|_{\mu_1=\dots=\mu_n=0}
    \end{equation}
    where $\chi_n^{A/Q}(\{\mu_i\}) = \psi_n^{A/Q}(\{\lambda_i\})$ is the scaled CGF for the position, see Eq.~\eqref{smeq:def_chi}. The change of variables is $\lambda_i = \mu_i + \dots + \mu_n$. Eq.~\eqref{smeq:PtnScaled} can be written in terms of $\mu_i$ as
    \begin{align} \label{smeq:PtnScaled2}
        \tilde P_n(\{\mu_i, \tau_i\}|z_0) &\sim 
        \int_{-\infty}^\infty \prod_{i=1}^n dz_i K_{\tau_i}(z_i, z_{i-1}) e^{\frac{\nu_0-\nu_i}{2}\mu_i}.
    \end{align}
    The derivative with respect to $\mu_i$ is non-zero only if $\nu_i=-\nu_0$ for $i=1,\dots, n$. The annealed cumulants of even order are easily obtained from Eqs.~\eqref{smeq:psiAscaled}, \eqref{smeq:defKappa} and \eqref{smeq:PtnScaled2} as
    \begin{align} \label{smeq:kappa2nAexpr}
        \kappa_{2n}^A(\{t_i\}) &\sim 2\sqrt{2} \int_{-\infty}^0 dz_0 P_{2n}^+(z_0, \{t_i\}), \\ 
        \label{smeq:PnP}
        P_n^+(z_0, \{t_i\}) &= \int_0^\infty \prod_{i=1}^n dz_i K_{t_i-t_{i-1}}(z_i, z_{i-1}).
    \end{align}
    We used the symmetry between $z_0<0$ and $z_0>0$. This corresponds to Eqs. (3) and (4) in the main text. The intepretation of this result in terms of a conditional probability of a Brownian walker is provided in Fig.~1c of the main text. 
    
    The expressions of the quenched cumulants are more complicated since their computation requires derivating the logarithm in Eq.~\eqref{smeq:psiQscaled}. They can still be expressed in terms of $P^+$. At orders $n=2$ and $n=4$:
    \begin{align} \label{smeq:exprKappa2Q}
        \kappa_{2}^Q(t_1, t_2) &\sim 2\sqrt{2} \int_{-\infty}^0 dz \left[P_2^+(z, t_1, t_2) - P_1^+(z, t_1)P_1^+(z, t_2) \right], 
    \end{align}
    \begin{multline}\label{smeq:exprKappa4Q}
        \kappa_{4}^Q(t_1, t_2, t_3, t_4) \sim 2\sqrt{2} \int_{-\infty}^0 dz \bigg[P_4^+(z, t_1, t_2, t_3, t_4) -\sum_{(ijk), l}P_3^+(z, t_i, t_j, t_k)P_1^+(z, t_l) 
        - \sum_{(ij),(kl)} P_2^+(z, t_i, t_j)P_2^+(z, t_k, t_l)\\+2\sum_{(ij),k,l}P_2^+(z, t_i, t_j)P_1^+(z, t_k)P_1^+(z, t_l) 
        - 6P_1^+(z, t_1)P_1^+(z, t_2) P_1^+(z, t_3)P_1^+(z, t_4)\bigg].
    \end{multline}
    In all sums, $\{i,j,k,l\} = \{1, 2, 3, 4\}$. The sum `$(ijk),l$' has 4 terms, the sum `$(ij),(kl)$' has 3 terms, and the sum `$(ij),k,l$' has 6 terms.

    \subsection{Explicit conditional probabilities}
    We now give the explicit expressions of the conditional probabilities $P_1^+, P_2^+$ and $P_3^+$. There is no simple expression for $P_4^+$ (unless one continues the series of integrals of a Gaussian: $\erfc(\bullet), T(\bullet,\bullet), S(\bullet,\bullet,\bullet), \dots$) but there is one for its integral over a half-space.
    Since $P_n^+$ is meant to be integrated over negative reals, we give the expressions for $(-z)$. We also consider  the time increments $\tau_i$ instead of the times $t_i$. Useful integrals are provided in section \ref{s:int}.
    \begin{align}
        P_1^+(-z, \tau_1) &= \frac{1}{2}\erfc\left(\frac{z}{\sqrt{\tau_1}}\right) \\
        P_2^+(-z, \tau_1, \tau_2) &= \frac{1}{4}\erfc\left(\frac{z}{\sqrt{\tau_1}}\right) + \frac{1}{4}\erfc\left(\frac{z}{\sqrt{\tau_1+\tau_2}}\right) -T\left(\frac{\sqrt{2}z}{\sqrt{\tau_1+\tau_2}}, \sqrt\frac{\tau_2}{\tau_1}\right)
        \label{smeq:P2p}\\
        P_3^+(-z, \tau_1, \tau_2, \tau_3) &= \frac{1}{8}\erfc\left(\frac{z}{\sqrt{\tau_1}}\right)\left[2-A\left(\frac{\tau_3}{\tau_2}\right)
        \right]
        + \frac{1}{8}\erfc\left(\frac{z}{\sqrt{\tau_1+\tau_2}}\right) + \frac{1}{4}\erfc\left(\frac{z}{\sqrt{\tau_1+\tau_2+\tau_3}}\right) \nonumber \\
        & -\frac{1}{2}T\left(\frac{\sqrt{2}z}{\sqrt{\tau_1+\tau_2}}, \sqrt\frac{\tau_2}{\tau_1}\right)-\frac{1}{2}T\left(\frac{\sqrt{2}z}{\sqrt{\tau_1+\tau_2+\tau_3}}, \sqrt\frac{\tau_3}{\tau_1+\tau_2}\right) \nonumber \\
        &-S\left(-\frac{\sqrt{2}z}{\sqrt{\tau_1+\tau_2+\tau_3}}, \sqrt\frac{\tau_3}{\tau_1+\tau_2}, \sqrt\frac{\tau_2(\tau_1+\tau_2+\tau_3)}{\tau_1\tau_3}\right)
    \end{align}
    The $\erf$, Owen-$T$ and Owen-$S$ functions are defined as succesive integrals of the standard Gaussian~\cite{Owen1980_SM,Brychkov2016_SM},
    \begin{align}
        \frac{1}{2}\erf\left(\frac{h}{\sqrt{2}}\right) &= \int_{0}^h \frac{e^{-\frac{x^2}{2}}}{\sqrt{2\pi}}  dx, \\
        T(h, a) &= \int_{-h}^\infty \frac{1}{2}\erf\left(\frac{ax}{\sqrt{2}}\right) \frac{e^{-\frac{x^2}{2}}}{\sqrt{2\pi}}  dx,\\
        S(h, a, b) &= \int_{-\infty}^h T(ax, b)  \frac{e^{-\frac{x^2}{2}}}{\sqrt{2\pi}} dx,
    \end{align}
    and $\erfc(h) = 1-\erf(h)$. As in the main text, we used the short notation $A(u) = \frac{2}{\pi}\atan\sqrt{u}$.
    
    \subsection{Results at large times}
    The Mathematica notebooks related to the computation of the cumulants are provided in~\cite{Poncet_TracerDenseSingleFile_2025_2_SM}.
    
    \subsubsection{Annealed cumulants}
    As explained in the main text, a striking consequence of Eqs.~\eqref{smeq:kappa2nAexpr}-\eqref{smeq:PnP} is that all the annealed cumulants of even order involving $k$ times, $\kappa^A_{\mathrm{even}, k}$, are equal. We give their expressions up to order $k=4$, using the integrals of section~\ref{s:int}.
    \begin{align}
    \label{smeq:kappaA1e}
        \kappa^A_{\mathrm{even}, 1}(t) &\sim 2\sqrt{2}\int_{-\infty}^0 dz\, P_1^+(z, t) = \sqrt\frac{2t}{\pi}, \\
        \kappa^A_{\mathrm{even}, 2}(t_1, t_2) &\sim 2\sqrt{2}\int_{-\infty}^0 dz\, P_2^+(z, t_1, t_2) = \frac{1}{\sqrt{2\pi}}\left(\sqrt{t_1} +\sqrt{t_2} - \sqrt{t_2-t_1}\right), \\
        \kappa_{\mathrm{even}, 3}^A(t_1, t_2, t_3) &\sim \frac{1}{2\sqrt{2}}\left[(1+a_{123})\sqrt{t_1} + \sqrt{t_2} + (1+b_{0123})\sqrt{t_3} - \sqrt{t_2-t_1} - \sqrt{t_3-t_1} - a_{012} \sqrt{t_3-t_2}\right], \\
        \kappa_{\mathrm{even}, 4}^A(t_1, t_2, t_3, t_4) &\sim \frac{1}{4\sqrt{2}}\bigg[(1+a_{123}+a_{124}+a_{134})\sqrt{t_1} + (1+a_{234})\sqrt{t_2} + (1+b_{0123}) \sqrt{t_3}\nonumber  \\ 
        &+ (1+b_{0124}+b_{0134}+b_{0234})\sqrt{t_4} -
        (1+a_{234})\sqrt{t_2-t_1} - \sqrt{t_3-t_1} - (1+b_{1234})\sqrt{t_4-t_1} \nonumber \\
        & - a_{012}\sqrt{t_3-t_2} - a_{012}\sqrt{t_4-t_2}- (a_{013}+a_{023}-a_{123})\sqrt{t_4-t_3}\bigg].
    \label{smeq:kappaA4e}
    \end{align}
    We defined
    \begin{align}
        A(u) &= \frac{2}{\pi}\atan\sqrt{u}, \label{eq:defA} &
        a_{ijk}&=A\left(\frac{t_j-t_i}{t_k-t_j}\right), &
        b_{ijkl}&=A\left(\frac{(t_j-t_i)(t_l-t_k)}{(t_l-t_i)(t_k-t_j)}\right).
    \end{align}
    Comments about these expressions are provided in the main text.

    \subsubsection{Quenched cumulants}
    The quenched cumulants of order $2$ and $4$ can be computed from Eqs.~\eqref{smeq:exprKappa2Q}-\eqref{smeq:exprKappa4Q}, using the integrals of section~\ref{s:int}.
    \begin{align}
    \label{smeq:kappa2Q}
        \kappa_2^Q(t_1, t_2) &\sim \frac{1}{\sqrt{2\pi}}\left(\sqrt{t_1+t_2} - \sqrt{t_2-t_1}\right), \\
    \label{smeq:kappa4Q}
        \kappa_4^Q(t_1, t_2, t_3, t_4) &\sim \frac{1}{4\sqrt{2\pi}}\sum_{1\leq i < j \leq 4}\left( D^{(4)}_{ij} \sqrt{t_j-t_i} + E^{(4)}_{ij} \sqrt{t_i+t_j} \right).
    \end{align}
    The coefficients $D$ and $E$ read
    \begin{align}
		D^{(4)}_{ij} &= a'_{ikl}-2c_{ikl} + \tilde D^{(4)}_{ij}, &
		E^{(4)}_{ij}=-2d_{ijkl}+6h_{ijkl} + \tilde E^{(4)}_{ij},
	\end{align}
	where $k, l$ are the two indices in $\{1, 2, 3, 4\}$ that are different from $i, j$ and such that $k<l$ (for instance if $i=1, j=3$ then $k=2, l=4$). The reduced coefficients $\tilde D^{(4)}_{ij}$ and $\tilde E^{(4)}_{ij}$ are
	\begin{align}
		\tilde D^{(4)}_{12} &= -a_{234}, &
		\tilde D^{(4)}_{13} &=  0, &
		\tilde D^{(4)}_{14} &=  -b_{1234}, \\
		\tilde D^{(4)}_{23} &=  1-a'_{421},&
		\tilde D^{(4)}_{24} &=  1-a'_{312},  &
		\tilde D^{(4)}_{34} &=  1+a_{123}-a'_{123}-a'_{213}, \\
		\tilde E^{(4)}_{12} &=  a_{134}+a_{234},&
		\tilde E^{(4)}_{13} &=  a_{124}-2g_{1324},&
		\tilde E^{(4)}_{14} &=  a_{123}+e'_{1423},\\
		\tilde E^{(4)}_{23} &=  -2g_{2314}-2g_{3214}, &
		\tilde E^{(4)}_{24} &=  e''_{2413}-2g_{4213},&
		\tilde E^{(4)}_{34} &=  e''_{3412}+e''_{4312}.
	\end{align}
	We needed to introduce the quantities $a_{ijk}, a'_{ijk}, b_{ijkl}, c_{ijkl}, d_{ijkl}, e_{ijkl}, e'_{ijkl}, f_{ijkl}, g_{ijkl}, h_{ijkl}$ are defined in terms of the function $A$ given by Eq.~\eqref{eq:defA}.
	\begin{align} 
		a_{ijk} &= A\left(\frac{t_j-t_i}{t_k-t_j}\right) &
		a'_{ijk} &= A\left(\frac{t_i+t_j}{t_k-t_j}\right)  \\
		b_{ijkl} &= A\left(\frac{(t_j-t_i)(t_l-t_k)}{(t_k-t_j)(t_l-t_i)}\right) &
		c_{ijk} &= A\left(\frac{t_i^2}{t_it_j+t_it_k+t_jt_k}\right) \\
		d_{ijkl} &= A\left(\frac{t_it_j+t_it_k+t_jt_k}{(t_i+t_j)(t_l-t_k)}\right)&
		e_{ijkl} &= A\left(\frac{(t_j-t_l)(t_i+t_k)}{(t_l-t_k)(t_i+t_j)}\right)\\
		e'_{ijkl} &= e_{ijkl}-2g_{ijkl}-2g_{ijlk} &
		e''_{ijkl} &= e'_{ijkl}+f_{ijkl} \\
		f_{ijkl} &= A\left(\frac{(t_k-t_i)(t_l-t_j)}{(t_i+t_j)(t_k+t_l)}\right) &
		g_{ijkl} &= A\left(\frac{t_i^2(t_j-t_k)}{(t_i+t_j)(t_it_k+t_it_l+t_kt_l)}\right) \\
		h_{ijkl} &= A\left(\frac{t_i^2t_j^2}{(t_i+t_j)[t_it_j(t_k+t_l) + t_kt_l(t_i+t_j)]}\right)
	\end{align}

    From these complicated expressions, one can deduce the cumulants $\kappa^Q_{(1222)}, \kappa^Q_{(1122)}, \kappa^Q_{(1112)}$ and $\kappa^Q_{(1223)}$ studied in the main text.

    \section{Two-point covariance and step initial density profile}
    \subsection{Two-point covariance}
    As a first extension, we consider two tracers $A$ and $B$ in the SEP. Their positions are denoted $X_A(t)$ and $X_B(t)$, with $X_A(0)=0, X_B(0) = L$. The single-time annealed CGF was computed in Ref.~\cite{Poncet2018_SM}. Here we focus on the two-time CGF
    \begin{equation}
        \Psi(\mu_1, \mu_2, t_1, t_2) =\lim_{\rho\to 1} \frac{\langle e^{\mu_1 X_A(t_1) + \mu_2 X_B(t_2) }\rangle}{1-\rho},
    \end{equation}
    where we restrict ourselves to probing particle $A$ at time $t_1$ and particle $B$ at time $t_2\geq t_1$.

    Once again, we look at the problem with a single vacancy and delimitate three zones: $\mathcal{A} = \{Z<0\}, \mathcal{B} = \{0<Z<L\}$ and $\mathcal{C} = \{Z>L\}$. In similarity with Eqs.~\eqref{eq:psina_hd}-\eqref{eq:psinq_hd}, the annealed and quenched dense CGFs are
    \begin{align}
	   \Psi^{A}(\{\mu_i, \tau_i\})  &= \sum_{Z\in\mathcal{A}} \left[ \tilde P_\mathcal{A}(\{\mu_i, \tau_i\})|Z) - 1 \right] + \sum_{Z\in\mathcal{B}} \left[ \tilde P_\mathcal{B}(\{\mu_i, \tau_i\})|Z) - 1 \right] + \sum_{Z\in\mathcal{C}} \left[ \tilde P_\mathcal{C}(\{\mu_i, \tau_i\})|Z) - 1 \right], \\
       \Psi^{Q}(\{\mu_i, \tau_i\})  &= \sum_{Z\in\mathcal{A}} \log \tilde P_\mathcal{A}(\{\mu_i, \tau_i\})|Z)+\sum_{Z\in\mathcal{B}} \log \tilde P_\mathcal{B}(\{\mu_i, \tau_i\})|Z)+\sum_{Z\in\mathcal{C}} \log \tilde P_\mathcal{C}(\{\mu_i, \tau_i\})|Z). \label{smeq:PsiQ2times} 
    \end{align}
    $\tilde P_\mathcal{A}$ encodes the contribution from a vacancy starting from $Z\in\mathcal{A}$, and so on.

    Instead of doing the full decomposition as in section~\ref{ss:prop}, we realize that in the single-vacancy problem the tracers can be displacements in $\{0,-1,+1\}$ only, leading to generating functions having factors $e^{\pm\mu_i}$. These factors can be obtained from the zones ($\mcA,\mcB,\mcC$) in which the vacancy is at times $t=0,t_1,t_2$. In details, we obtain
    \begin{align}
        \tilde P_\mcA(Z) &=  P^\mcA_{Z\to \mcA\to \mcA\mcB} + e^{-\mu_1}P^\mcA_{Z\to \mcB\mcC \to \mcA\mcB} + e^{-\mu_2} P^\mcA_{Z\to \mcA\to \mcB} + e^{-\mu_1-\mu_2} P^\mcA_{Z\to \mcB\mcC\to \mcC}, \\
        \tilde P_\mcB(Z) &=  P^\mcB_{Z\to \mcB\mcC\to \mcA\mcB} + e^{\mu_1}P^\mcB_{Z\to \mcA\to \mcA\mcB} + e^{-\mu_2} P^\mcB_{Z\to \mcB\mcC\to \mcC} + e^{\mu_1-\mu_2} P^\mcB_{Z\to \mcA\to \mcC}, \\ 
        \tilde P_\mcC(Z) &=  P^\mcC_{Z\to \mcB\mcC\to C} + e^{\mu_1}P^\mcC_{Z\to A\to C} + e^{\mu_2} P^\mcC_{Z\to \mcB\mcC\to \mcA\mcB} + e^{\mu_1+\mu_2} P^\mcC_{Z\to \mcA\to\mcA\mcB},
    \end{align}
    where $P^\mcA_{Z\to \mcA\to \mcA\mcB}$ is the probability that the vacancy is in zone $\mcA$ at time $t_1$ and in zone $\mcA$ or $\mcB$ at time $t_2$ knowing that it started from $Z\in\mcA$, and so on for the other probabilities. Collecting the terms involving $\mu_1$ and $\mu_2$, the annealed covariance reads
    \begin{align}
        \lim_{\rho\to 1} \frac{\langle X_A(t_1)X_B(t_2)\rangle_c^A}{1-\rho} = P_{\mcA\to \mcB\mcC\to \mcC}- P_{\mcB\to \mcA\to \mcC} + P_{\mcC\to \mcA\to\mcA\mcB} 
    \end{align}
    where $P_{\mcA\to \mcB\mcC\to \mcC} = \sum_{Z\in\mcA} P^\mcA_{Z\to \mcA\to \mcA\mcB}$.

    In the limit of large times $t_i\to\infty$, with $L/\sqrt{t_i} = \text{const}$, the sums becomes integrals, and the propagator of the vacancy is the Gaussian kernel $K_\tau(z_1, z_0)$. The three probabilities are therefore written as
    \begin{align}
		P_{A\to BC\to C} &\sim \sqrt{2}\int_{-\infty}^0 dz_0 \int_{0}^{\infty} dz_1  \int_{L}^{\infty} dz_2 K_{\tau_1}(z_1-z_0) K_{\tau_2}(z_2-z_1), \\
		P_{B\to A\to C} &\sim \sqrt{2}\int_0^{L} dz_0\,  \int_{-\infty}^{0} dz_1  \int_{L}^{\infty} dz_2 K_{\tau_1}(z_1-z_0) K_{\tau_2}(z_2-z_1), \\
		P_{C\to A\to AB}  &\sim \sqrt{2}\int_L^{\infty} dz_0   \int_{-\infty}^{0} dz_1  \int_{-\infty}^{L} dz_2 K_{\tau_1}(z_1-z_0) K_{\tau_2}(z_2-z_1),
	\end{align}
    where the boundaries of the integrals correspond to the zones ($\mcA,\mcB,\mcC$). The explicit computation of these integrals (see section \ref{s:int}) lead to
    \begin{align} \label{smeq:covarA}
        \lim_{\rho\to 1} \frac{\langle X_A(t_1)X_B(t_2)\rangle_c^A}{1-\rho} \sim \frac{1}{\sqrt{2\pi}}\left[g_L(t_1) + g_L(t_2) - g_L(|t_1-t_2|)\right]
    \end{align}
    with
    \begin{equation}
        g_L(t) =  \sqrt{t}\left[e^{-\frac{L^2}{2t}} - \sqrt{\pi} \frac{L}{\sqrt{2t}} \erfc \frac{L}{\sqrt{2t}}\right].
    \end{equation}
    This result is consistent with the Edwards-Wilkinson interface with equilibrium initial conditions. 
    We note that beyond this Gaussian order, Eq.~\eqref{smeq:covarA} actually holds for any even cumulant of the form $\langle X_A(t_1)^nX_B(t_2)^m\rangle_c^A$.

    The computation of the quenched covariance is a little more involved.
    By derivating Eq.~\eqref{smeq:PsiQ2times} with respect to $\mu_1$ and $\mu_2$, one obtains
    \begin{multline}
        \lim_{\rho\to 1} \frac{\langle X_A(t_1)X_B(t_2)\rangle_c^Q}{1-\rho} \sim \int_\mcA dz \left(P^\mcA_{z\to \mcB\mcC\to \mcC} - P^\mcA_{z\to \mcB\mcC\to \mcA\mcB\mcC} P^\mcA_{z\to \mcA\mcB\mcC\to \mcC}\right)
        + \int_\mcB dz \left(-P^\mcB_{z\to \mcA\to \mcC} + P^\mcB_{z\to \mcA\to \mcA\mcB\mcC} P^\mcB_{z\to \mcA\mcB\mcC\to \mcC}\right) \\
        + \int_\mcC dz \left( P^\mcC_{z\to \mcA\to \mcA\mcB} - P^\mcC_{z\to \mcA\to \mcA\mcB\mcC}P^C_{z\to \mcA\mcB\mcC\to \mcA\mcB}\right).
    \end{multline}
    All the integrals can be evaluated explicitly, leading to
    \begin{equation}
        \lim_{\rho\to 1} \frac{\langle X_A(t_1)X_B(t_2)\rangle_c^Q}{1-\rho}\sim \frac{1}{\sqrt{2\pi}}\left[g_L(t_1+t_2)- g_L(|t_2-t_1|)\right].
    \end{equation}
    This result is in line with the Edwards-Wilkinson interface with flat initial conditions. 

    The reader may remark that the framework can be extended to an arbitrary number of tracers and an arbitrary number of times. We were however limited by the absence of explicit expressions for the integrals (to the best of our knowledge), beyond the ones involved in the covariances computed above.

    \subsection{Step initial condition}
    We define a step initial condition as follow. We consider a system of $2N+1$ sites (at position $r$ with $-N\leq r\leq N$), with the tracer at the origin. The number of occupied positive sites is $N-M_+$ and the number of occupied negative sites is $N-M_-$. The density in front of / behind the tracer is $\rho_\pm = 1-M_\pm/N$.
    In this setup, the large density limit Eq.~\eqref{smeq:PnHigh} becomes
    \begin{align}
        \mathcal{\tilde P}^{(n)}(\{\lambda_i, \tau_i\}|\{Z_j^+\},\{Z_j^-\}) &\equi{\rho_\pm\to 1} \prod_{j=1}^{M_+} \tilde  P^{(n)}(\{\lambda_i, \tau_i\}|Z_j^+)
        \prod_{j=1}^{M_-} \tilde  P^{(n)}(\{\lambda_i, \tau_i\}|Z_j^-),
    \end{align}
    where $Z_j^+$ are the positions of the positive vacancies, and similarly for $Z_j^-$. 
    The CGFs, Eqs.~\eqref{eq:psina_hd}-\eqref{eq:psinq_hd}, have the following high-density expression:
	\begin{align}
	   \tilde \psi^{A}_n(\{\lambda_i, \tau_i\}) &\equi{\rho_\pm\to 1} (1-\rho_+)\sum_{Z>0} \left[ \tilde P^{(n)}(\{\lambda_i, \tau_i\}|Z) - 1 \right]+(1-\rho_-)\sum_{Z<0} \left[ \tilde P^{(n)}(\{\lambda_i, \tau_i\}|Z) - 1 \right], \\
	   \tilde \psi^{Q}_n(\{\lambda_i, \tau_i\}) &\equi{\rho_\pm\to 1} (1-\rho_+)\sum_{Z>0} \log \tilde P^{(n)}(\{\lambda_i, \tau_i\}|Z) +(1-\rho_-)\sum_{Z<0} \log\tilde P^{(n)}(\{\lambda_i, \tau_i\}|Z).
	\end{align}
    It is convenient to define the average density $\rho=(\rho_++\rho_-)/2$ and the step parameter $\sigma = (\rho_--\rho_+)/[2(1-\rho)]$. The high-density CGFs now read
    \begin{align}
        \psi^{A}_n(\{\lambda_i, \tau_i\}) &= \sum_{Z\neq 0} (1+\nu \sigma)  \left[ \tilde P^{(n)}(\{\lambda_i, \tau_i\}|Z) - 1 \right], \\
        \psi^{Q}_n(\{\lambda_i, \tau_i\}) &= \sum_{Z\neq 0} (1+\nu \sigma)  \log\tilde P^{(n)}(\{\lambda_i, \tau_i\}|Z),
    \end{align}
    with $\nu=\mathrm{sign}(Z)$. The annealed cumulants are expressed in terms of $P_n^+$ (Eq.~\eqref{smeq:PnP}) as
    \begin{align} \label{smeq:kappa2nAexpr2}
        \kappa_{2n}^A(\{t_i\}) &\sim 2\sqrt{2} \int_{-\infty}^0 dz_0 P_{2n}^+(z_0, \{t_i\}), &
        \kappa_{2n+1}^A(\{t_i\}) &\sim 2\sqrt{2}\sigma \int_{-\infty}^0 dz_0 P_{2n+1}^+(z_0, \{t_i\}).
    \end{align}
    All cumulants of even order involving $k$ times, $\kappa^A_{\mathrm{even}, k}$, are equal and all cumulants of odd order, $\kappa^A_{\mathrm{odd}, k}$ are also equal. They are linked by the relation $\kappa^A_{\mathrm{odd}, k} = \sigma\kappa^A_{\mathrm{even}, k}$. Up to order $k=4$, the expressions are given by Eqs.~\eqref{smeq:kappaA1e}-\eqref{smeq:kappaA1e}.

    As for the quenched cumulants, the even ones $\kappa_{2n}^Q$ are unchanged compared with the case of uniform initial conditions, see Eqs.~\eqref{smeq:kappa2Q}-\eqref{smeq:kappa4Q}. The odd ones have new non-zero expressions. The displacement is the same as for annealed initial conditions $\kappa_1^Q \sim \kappa_1^A \sim \sigma\sqrt{2t/\pi}$. The third cumulant is
    \begin{align}
        \kappa_3^Q(t_1, t_2, t_3) &\sim 2\sqrt{2}\sigma \int_{-\infty}^0 dz \bigg[P_3^+(z, t_1, t_2, t_3) - \sum_{(ij), k} P_2^+(z, t_i, t_j) P_1^+(z, t_k) + 2P_1^+(z, t_1)P_1^+(z, t_2)P_1^+(z, t_3)\bigg] \\
        &\sim \frac{\sigma}{2\sqrt{2\pi}} \left[\sum_{0\leq i<j\leq 3} D_{ij}^{(3)} \sqrt{t_j-t_i} + \sum_{1\leq i<j\leq 3} E_{ij}^{(3)} \sqrt{t_i+t_j}\right] 
    \end{align}
    with the following coefficients.
    \begin{align}
        D_{01}^{(3)} &= A\left(\frac{t_3-t_2}{t_2}\right) - A\left(\frac{t_3-t_2}{t_2-t_1}\right) &
        D_{02}^{(3)} &= -A\left(\frac{t_1}{t_3-t_1}\right) \\
        D_{03}^{(3)} &= A\left(\frac{t_2-t_1}{t_1}\right) - A\left(\frac{t_3(t_2-t_1)}{t_1(t_3-t_2)}\right) &
        D_{12}^{(3)} &= -A\left(\frac{t_1}{t_3}\right)  \\
        D_{13}^{(3)} &= -A\left(\frac{t_1}{t_2}\right) &
        D_{23}^{(3)} &= A\left(\frac{t_2-t_1}{t_1}\right) - A\left(\frac{t_2}{t_1}\right) \\
        E_{12}^{(3)} &= 2A\left(\frac{t_1t_2}{t_3(t_1+t_2)}\right) &
        E_{13}^{(3)} &= 2A\left(\frac{t_1t_2}{t_2(t_1+t_3)}\right) - A\left(\frac{t_1(t_3-t_2)}{t_2(t_1+t_3)}\right) \\
        E_{23}^{(3)} &= 2A\left(\frac{t_2t_3}{t_1(t_2+t_3)}\right) - A\left(\frac{t_3(t_2-t_1)}{t_1(t_2+t_3)}\right) - A\left(\frac{t_2(t_3-t_1)}{t_1(t_2+t_3)}\right)
    \end{align}

    \section{Biased tracer}
    We now study the case of a biased tracer. It has the same exponential clock as the other particles ($\Xi(t)=e^{-t}$) but when the clock rings, it has a probability $p_+=(1+s)/2$ to attempt a jump to the right, and $p_-=(1-s)/2$ to try to jump to the left, with $-1\leq s \leq 1$.
    
    \subsection{Expression of the propagators}
    The random walk performed by a single vacancy is no longer translationaly invariant. The jumps next to the tracer acquire a special status. The density of first passage to the origin starting from the site on the right of the tracer, $f_1(t)$, can be obtained by considering that the first move is either made by the tracer or by the particle on site $2$, see Ref.~\cite{Poncet2022_SM}. 
    One should be careful that the rate at which such a move happens is $(p_++1/2)$, and not $1$. The decomposition is:
    \begin{equation}
        f_1(t) = p_+ e^{-(p_+ + \frac{1}{2})t} + \int_0^\infty dt_1 dt_2 dt_3 \delta(t-t_1-t_2-t_3) \frac{1}{2} e^{-(p_+ + \frac{1}{2})t_1} f_1^U(t_2) f_1(t_3),
    \end{equation}
    where $\hat f_1^U(u) = \alpha$ is the quantity computed before for an unbiased walk. In the Laplace domain, we obtain
    \begin{equation}
        \hat f_1(u) = \frac{2p_+\alpha}{1+\alpha(2p_+-1)} = \frac{(1+s)\alpha}{1+s\alpha}.
    \end{equation}
    The partitioning to get $\hat f_Z(u)$ (first passage density starting from $Z$) is the same as in the unbiased case:
    \begin{align}
        \hat f_Z(u) = \frac{(1+\nu s)}{1+\nu s \alpha} \alpha^Z
    \end{align}
    where $\nu$ is the sign of $Z$.

    We now compute the probability $\Pi_{Z_1,Z_0}(t)$ that the walker goes from $Z_0$ to $Z_1$ during $t$ without touching the origin. For simplicity, we focus on the case $Z_0,Z_1\geq 2$. The two decompositions, when the walker starts for $1$ or from $Z_0\geq 2$, are
    \begin{align}
        \Pi_{Z_1, 1}(t) &= \int_0^t dt_0 \frac{1}{2} e^{-(p_++1/2)t_0} \Pi_{Z_1, 2}(t-t_0), \\
        \Pi_{Z_1, Z_0}(t) &= \Pi^U_{Z_1-1, Z_0-1}(t) 
    + \int_0^t dt_0 f^U_{Z_0-1}(t_0) \Pi_{Z_1, 1}(t-t_0),
    \end{align}
    where $f^U$ and $\Pi^U$ refer to the quantities in the unbiased case. Combining the two equations in the Laplace domain, we obtain successively,
    \begin{align}
    \hat \Pi_{Z_1, 1}(u) &= \frac{\hat\Pi^\mathrm{UB}_{Z_1-1, 1}(u)}{2(u+p_+) + 1 - \alpha} 
    = \frac{2\alpha^{Z_1}}{1+\alpha s}, \\
    \hat\Pi_{Z_1, Z_0}(u) &= \frac{1}{u} \frac{1-\alpha}{1+\alpha}
    \left(\alpha^{|Z_1-Z_0|}-\frac{\alpha + s}{1+\alpha s}\alpha^{|Z_1| + |Z_0|-1}\right).
    \end{align}
    The last equation remains valid for $Z_0, Z_1>0$. For $Z_0, Z_1<0$, $s$ should be changed into $-s$. And if $Z_0$ and $Z_1$ have opposite signs, $\hat \Pi_{Z_1, Z_0}=0$ by definition.

    We can now compute $\hat{\tilde P}^\dagger(\lambda, u, Z_1|Z_0)$ from Eq.~\eqref{eq:Pdagger2}:
    \begin{equation}
    \label{eq:PdaggerBiased}
    \hat{\tilde P}^\dagger(\lambda, u, Z_1|Z_0) = \frac{1}{u} \frac{1-\alpha}{1+\alpha}
		\begin{cases}
			\alpha^{|Z_0-Z_1|} - \nu_0 s \alpha^{|Z_0|+|Z_1|-1}  & \text{ if } \nu_0 = \nu_1 \\
			(1+\nu_0s) \alpha^{|Z_0|+|Z_1|-1} e^{\nu_0\lambda} & \text{ if } \nu_0 = - \nu_1
		\end{cases},        
    \end{equation}
    with $\nu_i$ the sign of $Z_i$.
    
    \subsection{Cumulant-generating functions at large times}
    The large-time limit ($u\to 0$, $Z_i\sqrt{u}=\text{const}$, or equivalently $t\to \infty$, $Z_i/\sqrt{t}=\text{const}$) of Eq.~\eqref{eq:PdaggerBiased} is
    \begin{align}
        \hat{\tilde P}^\dagger(\lambda, u, Z_1| Z_0) &\equi{u\to 0} \frac{1}{\sqrt{2u}}\left[ e^{-|Z_0-Z_1|\sqrt{2u}} - \nu_1 s e^{-(|Z_0|+|Z_1|)\sqrt{2u}} \right]e^{\frac{\nu_0-\nu_1}{2}\lambda}, \\
        \tilde P^\dagger(\lambda, t, Z_1| Z_0) &\equi{t\to \infty} \frac{1}{\sqrt{2\pi t}} \left[ e^{-\frac{(Z_0-Z_1)^2}{2t}} - \nu_1 s e^{-\frac{(|Z_0|+|Z_1|)^2}{2t}}\right] e^{\frac{\nu_0-\nu_1}{2}\lambda}.
    \end{align}
    As a consequence, $\tilde P_n(\{\lambda_i, \tau_i\}|z_0)$, see Eq.~\eqref{smeq:PtnScaled}, reads
     \begin{align}
    \label{smeq:PtnScaledB}
        \tilde P_n(\{\lambda_i, \tau_i\}|z_0) &\sim 
        \int_{-\infty}^\infty \prod_{i=1}^n dz_i K_{\tau_i, s}(z_i, z_{i-1}) e^{\frac{\nu_{i-1}-\nu_i}{2}\lambda_i},
    \end{align}
    with the biased kernel
    \begin{equation}
        K_{\tau,s}(z', z) = \frac{1}{\sqrt{\pi\tau}} \left[ e^{-\frac{(z'-z)^2}{\tau}} -\nu' s e^{-\frac{(|z'|+|z|)^2}{\tau}}\right].
    \end{equation}
    As in the unbiased case, the annealed and quenched CGF, Eqs.~\eqref{eq:psina_hd}-\eqref{eq:psinq_hd}, are computed as
    \begin{align} \label{smeq:psiAscaledB}
        \psi_n^A(\{\lambda_i, \tau_i\}) &\sim \sqrt{2} \int_{-\infty}^\infty dz_0\left[\tilde P_n(\{\lambda_i, \tau_i\}|z_0) - 1\right], \\
        \label{smeq:psiQscaledB}
        \psi_n^Q(\{\lambda_i, \tau_i\})&\sim \sqrt{2} \int_{-\infty}^\infty dz_0\log \tilde P_n(\{\lambda_i, \tau_i\}|z_0).
    \end{align}

    \subsection{Cumulants at large times from a Brownian motion}
    We start from Eq.~\eqref{smeq:PtnScaledB}, make the change of variables $\mu_i=\lambda_i+\dots+\lambda_n$ and consider the $n$-th derivative of $\psi_n^A$ from Eq.~\eqref{smeq:psiAscaledB}. We obtain the annealed cumulants as
    \begin{align}
        \kappa_n^A(t_1, \dots, t_n) &\sim \sqrt{2}\left\{ \int_0^\infty dz_0 \int_{-\infty}^0\prod_{i=1}^n dz_i K_{\tau_i, s}(z_i, z_{i-1}) + (-1)^n  \int_{-\infty}^0 dz_0 \int_0^\infty\prod_{i=1}^n dz_i K_{\tau_i, s}(z_i, z_{i-1}) \right\}, \\
        &\sim \sqrt{2} \int_{-\infty}^0 dz_0 \int_0^\infty\left(\prod_{i=1}^n dz_i\right) \left[ \prod_{i=1}^n K_{\tau_i, -s}(z_i, z_{i-1})+(-1)^n \prod_{i=1}^n K_{\tau_i, s}(z_i, z_{i-1})\right].
    \end{align}
    We used the symmetry relation $(z_i, s)\leftrightarrow (-z_i, -s)$.
    Assuming $z_0<0$ and $z_i>0$ ($i\geq 1$), we can recast the product of biased kernels in term of a sum of product of the unbiased kernel $K_\tau(z', z) = e^{-z^2/\tau}/\sqrt{\pi\tau}$:
    \begin{align}
        \prod_{i=1}^n K_{\tau_i, -s}(z_i, z_{i-1}) &= (1+s)K_{\tau_1}(z_1,z_0) \sum_{\epsilon_2, \dots,\epsilon_n=0}^1 \prod_{i=2}^n s^{\epsilon_i} K_{\tau_i }(z_i, [1-2\epsilon_i]z_{i-1}), \\
        \prod_{i=1}^n K_{\tau_i, s}(z_i, z_{i-1}) &= (1-s)K_{\tau_1}(z_1,z_0) \sum_{\epsilon_2, \dots,\epsilon_n=0}^1 \prod_{i=2}^n (-s)^{\epsilon_i} K_{\tau_i }(z_i, [1-2\epsilon_i]z_{i-1}).
    \end{align}
    Summing the two contributions, we obtain
    \begin{align}
        \kappa_n^A(t_1, \dots, t_n) 
        &\sim 2\sqrt{2} \int_{-\infty}^0 dz_0 \int_0^\infty\left(\prod_{i=1}^n dz_i\right) \sum_{\epsilon_2, \dots,\epsilon_n=0}^1  s^{\sum_i\epsilon_i} S_n(\underline{\epsilon})K_{\tau_1}(z_1,z_0) \prod_{i=1}^n K_{\tau_i }(z_i, [1-2\epsilon_i]z_{i-1}),
    \end{align}
    with
    \begin{align}
        S_n(\underline{\epsilon}) = \frac{1}{2}\left[(1+s) + (1-s)(-1)^{n+\sum_i\epsilon_i}\right] =
        \begin{cases}
            1 & \text{if $n$ is even and $\sum_i\epsilon_i$ is even} \\
            s & \text{if $n$ is even and $\sum_i\epsilon_i$ is odd} \\
            s & \text{if $n$ is odd and $\sum_i\epsilon_i$ is even} \\
            1 & \text{if $n$ is odd and $\sum_i\epsilon_i$ is odd}
        \end{cases}.
    \end{align}
    This can be recast as
    \begin{align} \label{smeq:kappa_n_s}
    \kappa_n^A(t_1, \dots, t_n) 
        &\sim 2\sqrt{2} \sum_{\epsilon_2, \dots, \epsilon_n} s^{M(\underline{\epsilon})} \int_{-\infty}^0 dz_0  P_n^{\underline{\epsilon}}(z_0;t_1,\dots, t_n), \\
        P_n^{\underline{\epsilon}}(z_0;t_1,\dots, t_n) &=
        \int_0^\infty \left(\prod_{i=1}^n dz_i\right) K_{\tau_1}(z_1,z_0) \prod_{i=1}^n K_{\tau_i }(z_i, [1-2\epsilon_i]z_{i-1})
    \end{align}
    where $M(\underline{\epsilon})$ is the next nearest integer from $\sum_i\epsilon_i$. As explained in the main text, $P_n^{\underline{\epsilon}}(z_0)$ has a simple interpretation: its is the probability that an unbiased Brownian walker starting from $z_0$ is at a positive position at $t_1$ and then changes sign between $t_{i-1}$ and $t_i$ according to $\epsilon_i$ for $i=2,\dots, n$ ($\epsilon_i = 1$: change of sign, $\epsilon_i=0$ no change of sign).
    
    Let us be explicit in the cases $n=1$ and $n=2$:
    \begin{align}
        \kappa_1^A(t) &\sim 2\sqrt{2} s \int_{-\infty}^0 dz_0 P_1^{+}(z_0, t), \\
        \kappa_2^A(t_1, t_2) &\sim 2\sqrt{2} \int_{-\infty}^0 dz_0 \left[P_2^{(\epsilon_2=0)}(z_0, t_1, t_2) + s^2P_2^{(\epsilon_2=1)}(z_0, t_1, t_2)\right],
    \end{align}
    where $P_2^{(\epsilon_2=0)} = P_2^+$ is given by Eq.~\eqref{smeq:P2p} and
    \begin{equation}
        P_2^{(\epsilon_2=1)}(-z, \tau_1, \tau_2) = \frac{1}{4}\erfc\left(\frac{z}{\sqrt{\tau_1}}\right) - \frac{1}{4}\erfc\left(\frac{z}{\sqrt{\tau_1+\tau_2}}\right) + T\left(\frac{\sqrt{2}z}{\sqrt{\tau_1+\tau_2}}, \sqrt\frac{\tau_2}{\tau_1}\right).
    \end{equation}

    The quenched cumulants are computed from Eqs.~\eqref{smeq:PtnScaledB} and \eqref{smeq:psiQscaledB} as
    \begin{equation}
        \kappa_n^Q(t_1, \dots, t_n) \sim \sqrt{2}  \int_{-\infty}^\infty dz_0 \frac{\partial^n}{\partial\mu_1\dots\partial\mu_n}\left(\log \int_{-\infty}^\infty \prod_{i=1}^n dz_i K_{\tau_i, s}(z_i, z_{i-1}) e^{\frac{\nu_{0}-\nu_i}{2}\mu_i} \right)_{\mu_1=\dots=\mu_n=0}.
    \end{equation}
    The orders $n=1$ and $n=2$ give $\kappa_1^Q(t) \sim \kappa_1^A(t)$ and
    \begin{equation}
        \kappa_2^Q(t_1, t_2) \sim 2\sqrt{2} \int_{-\infty}^0 dz_0 \left[P_2^{(\epsilon_2=0)}(z_0, t_1, t_2) + s^2P_2^{(\epsilon_2=1)}(z_0, t_1, t_2) - P_1^+(z_0, t_1)P_1^+(z_0, t_2)\right].
    \end{equation}
    
    \subsection{Results at large times}
    The Mathematica notebooks related to the computation of the cumulants are provided in~\cite{Poncet_TracerDenseSingleFile_2025_2_SM}.
    
    \subsubsection{Annealed cumulants}
    Looking at Eq.~\eqref{smeq:kappa_n_s}, one sees that all the even annealed cumulants involving $k$ times, $\kappa_{\mathrm{even}, k}$ are equal. The same goes for the odd cumulants, $\kappa_{\mathrm{odd}, k}$. The results can be summarized by the formulas
    \begin{align}
        \kappa_{\mathrm{even}, k}^A &\sim \frac{2^{2-k}}{\sqrt{2\pi}}\sum_{0\leq i < j \leq k} C^{(k, e)}_{ij} \sqrt{t_j-t_i}, \\
        \kappa_{\mathrm{odd}, k}^A &\sim s\frac{2^{2-k}}{\sqrt{2\pi}}\sum_{0\leq i < j \leq k} C^{(k, o)}_{ij} \sqrt{t_j-t_i}.
    \end{align}
    We give the values of the coefficients $C^{(k, x)}_{ij}$ below.
	\begin{align}
		C^{(1,e)}_{01} &= 1 & C^{(1,o)}_{01} &= 1 \\[0.2cm]
		C^{(2,e)}_{01} &= u_+ & C^{(2,e)}_{02} &= u_- & C^{(2,e)}_{12} &= -u_-\\
		C^{(2,o)}_{01} &= 2 & C^{(2,o)}_{02} &= 0 & C^{(2,o)}_{12} &= 0 \\[0.2cm]
		C^{(3,e)}_{01} &= 1+3s^2 +u_- a_{123} & C^{(3,e)}_{02} &= u_- & C^{(3,e)}_{03} &= u_-(b_{0123}+1) \\
		C^{(3,e)}_{12} &= -u_- & C^{(3,e)}_{13} &= -u_- & C^{(3,e)}_{23} &= -u_-a_{012} \\
		C^{(3,o)}_{01} &= 3+s^2 + u_-a_{123} & C^{(3,o)}_{02} &= u_- & C^{(3,o)}_{03} &= u_-(b_{0123}-1) \\
		C^{(3,o)}_{12} &= -u_- & C^{(3,o)}_{13} &= u_- & C^{(3,o)}_{23} &= -u_-a_{012}
	\end{align}
	\begin{align}
		\tilde C^{(4,e)}_{01} &= 8s^2 + u_-\left(u_- + u_+a_{123}+u_-a_{124}+u_+a_{134}\right) & 
		\tilde C^{(4,e)}_{02} &= u_-\left(u_+ + u_-a_{234}\right) \\
		\tilde C^{(4,e)}_{03} &= u_-\left(u_-+u_+b_{0123}\right)  &
		\tilde C^{(4,e)}_{04} &= u_-\left(u_+ + u_-b_{0124} + u_+ b_{0134} + u_- b_{0234}\right) \\
		\tilde C^{(4,e)}_{12} &= -u_-\left(u_+ + u_-a_{234}\right) &
		\tilde C^{(4,e)}_{13} &= -u_-^2 \\
		\tilde C^{(4,e)}_{14} &=  -u_-\left(u_+ + u_-b_{1234}\right) &
		\tilde C^{(4,e)}_{23} &=  -u_-u_+ a_{012} \\
		\tilde C^{(4,e)}_{24} &=  -u_-^2 a_{012} &
		\tilde C^{(4,e)}_{34} &=  -u_-\left(u_+ a_{013} + u_- a_{023} - u_- a_{123}\right) \\[0.2cm]
		\tilde C^{(4,o)}_{01} &= 4u_+ +2u_-\left(a_{123}+a_{134}\right) & 
		\tilde C^{(4,o)}_{02} &= 2u_- \\
		\tilde C^{(4,o)}_{03} &= 2u_-b_{0123}  &
		\tilde C^{(4,o)}_{04} &= 2u_-\left(b_{0134} - 1\right) \\
		\tilde C^{(4,o)}_{12} &= -2u_- &
		\tilde C^{(4,o)}_{13} &= 0 \\
		\tilde C^{(4,o)}_{14} &=  2u_- &
		\tilde C^{(4,o)}_{23} &=   -2u_- a_{012} \\
		\tilde C^{(4,o)}_{24} &=  0 &
		\tilde C^{(4,o)}_{34} &=  -2u_- a_{013}
	\end{align}
	We defined $u_+ = 1+s^2 = 8(p_+^2-p_-^2)$ and $u_- = 1-s^2 = 4p_+p_-$ and used the short notations of Eq.~\eqref{eq:defA}.
    
    \subsubsection{Quenched cumulants}
    The quenched cumulants, up to order $n=3$, are given by
    \begin{align}
         \kappa_1^Q &\sim s\sqrt\frac{2t}{\pi} \\
         \kappa_2^Q &\sim \frac{1}{\sqrt{2\pi}}\left[(1+s^2)\sqrt{t_1+t_2} - (1-s^2)\sqrt{t_2-t_1} -2s^2\sqrt{t_2}\right] \\
         \kappa_3^Q &\sim \frac{s}{2\sqrt{2\pi}}\left[\sum_{0\leq i<j\leq 3} D_{ij}^{(3)} \sqrt{t_j-t_i} + \sum_{1\leq i<j\leq 3} E_{ij}^{(3)} \sqrt{t_i+t_j}\right] 
     \end{align}
     The coefficients for $\kappa_3^Q$ are
     \begin{align}
         D_{01}^{(3)} &= u_-\left[a_{123}-a_{012}\right] &
         D_{02}^{(3)} &= -u_-a_{013} &
         D_{03}^{(3)} &= 4s^2 + u_-\left[b_{0123} - a_{012} \right] \\
         D_{12}^{(3)} &= -u_- A\left(\frac{t_1}{t_3}\right)&
         D_{13}^{(3)} &= u_-\left[2 - A\left(\frac{t_1}{t_2}\right)\right] &
         D_{23}^{(3)} &= u_-\left[A\left(\frac{t_1}{t_2}\right)-a_{012}\right] \\
         E_{12}^{(3)} &=  2(3+s^2)\left[1-n_{312}\right] &
         E_{13}^{(3)} &= 4 - u_- m_{123} -2(3+s^2) n_{213} &
         E_{23}^{(3)} &= u_-\left[2-m_{213}-m_{312}\right]-2(3+s^2) n_{123}
         \end{align}
         with  $u_\pm = 1\pm s^2$ and
         \begin{align}
             m_{ijk} &= A\left(\frac{t_i(t_k-t_j)}{t_j(t_i+t_k)}\right) &
             n_{ijk} &= A\left(\frac{t_it_j}{t_k(t_i+t_j)}\right).
         \end{align}

    With the help of Mathematica, we are also able to obtain an explicit expression for $\kappa_4^Q$. It takes the form 
    \begin{align}
         \kappa_4^Q &\sim \frac{s}{4\sqrt{2\pi}}\left[\sum_{0\leq i<j\leq 4} D_{ij}^{(4)} \sqrt{t_j-t_i} + \sum_{1\leq i<j\leq 4} E_{ij}^{(4)} \sqrt{t_i+t_j}\right].
     \end{align}
     The coefficients $D_{ij}^{(4)}$ and $E_{ij}^{(4)}$ are expressed in terms of the function $A$, but their expressions are too long to be reported here, see~\cite{Poncet_TracerDenseSingleFile_2025_2_SM}.

    \section{Useful formulas} \label{s:int}
    We give here a list of useful integrals, and symmetry relations involving Gaussians and integrals of Gaussians. Some integrals are known in the litterature \cite{Owen1980_SM,prudnikov1992integrals_SM,Brychkov2016_SM}. For some others, we are not aware of previous computations. Unless specified, we assume $\alpha, \beta,\gamma, a,b > 0$. 

      \subsection{Gaussian integrals}
    \begin{align}
        \int_0^\infty dz\, e^{-\alpha^2(z-x)^2} &= \frac{1}{2}\sqrt\frac{\pi}{\alpha}\erfc(-\alpha x) \\
        \int_{-\infty}^\infty dz\, \frac{e^{-\frac{(x-z)^2}{\tau}}}{\sqrt{\pi\tau}} \frac{e^{-\frac{(z-y)^2}{\eta}}}{\sqrt{\pi\eta}} &= \frac{e^{-\frac{(x-y)^2}{\tau+\eta}}}{\sqrt{\pi(\tau+\eta)}} \\
        \int_{0}^\infty dz \frac{e^{-\frac{(x-z)^2}{\tau}}}{\sqrt{\pi\tau}} \frac{e^{-\frac{(z-y)^2}{\eta}}}{\sqrt{\pi\eta}} &= \frac{e^{-\frac{(x-y)^2}{\tau+\eta}}}{\sqrt{\pi(\tau+\eta)}} \frac{1}{2} \left[1+\erf\left(\frac{\eta x + \tau y}{\sqrt{\tau\eta(\eta+\tau)}}\right)\right]
    \end{align}

    \subsection{Integrals involving erfc}
    \begin{align}
        \int_0^\infty dz\, \erfc(\alpha z) &= \frac{1}{\alpha\sqrt{\pi}} \\
        \int_0^\infty dz\, \erfc(\alpha z)\erfc(\beta z) &=  \frac{1}{\alpha\sqrt{\pi}} + \frac{1}{\beta\sqrt{\pi}} -\frac{\sqrt{\alpha^2+\beta^2}}{\alpha \beta\sqrt{\pi}} \\
        \int_0^\infty dz\, \prod_{i=1}^3 \erfc(\alpha_i z)  &= \sum_{i=1}^3 \frac{1}{\alpha_i\sqrt{\pi}} - \sum_{\sigma\in\mathcal{S}_3}f_3(\alpha_{\sigma_1}, \alpha_{\sigma_2}, \alpha_{\sigma_3})\\
        \int_0^\infty dz\, \prod_{i=1}^4 \erfc(\alpha_i z) &= \sum_{i=1}^4 \frac{1}{\alpha_i\sqrt{\pi}} + \sum_{\sigma\in\mathcal{S}_4}\left[\frac{1}{2}f_4(\alpha_{\sigma_1}, \alpha_{\sigma_2}, \alpha_{\sigma_3}, \alpha_{\sigma_4})-f_3(\alpha_{\sigma_1}, \alpha_{\sigma_2}, \alpha_{\sigma_3})\right] \\
        f_3(\alpha, \beta, \gamma) &= \frac{\sqrt{\alpha^2 +\beta^2}}{\pi^{3/2} \alpha\beta} \arctan\frac{\sqrt{\alpha^2+\beta^2}}{\gamma} \\
        f_4(\alpha, \beta, \gamma, \delta) &= \frac{\sqrt{\alpha^2 +\beta^2}}{\pi^{3/2} \alpha\beta} \arctan\frac{\sqrt{\alpha^2 +\beta^2}\sqrt{\alpha^2 +\beta^2+\gamma^2+\delta^2}}{\gamma\delta} \\
        \int_0^\infty dz\, e^{-\alpha^2(z-x)^2} \erfc(\beta z) &= \frac{1}{2} \sqrt{\pi \alpha} \left[\erfc\left(\frac{\alpha\beta}{\sqrt{\alpha^2+\beta^2}}x\right) -\erfc(\alpha x) + 4 T\left(\frac{\alpha\beta\sqrt{2}}{\sqrt{\alpha^2+\beta^2}}x, \frac{\alpha}{\beta}\right)\right] 
    \end{align}
    \begin{align} 
		\int_x^\infty \erfc(\alpha z) dz &= \frac{1}{a\sqrt\pi} g(\alpha x)\\
		\int_0^\infty \erfc[\alpha (x+z)] \erfc(\beta z) dz &= \frac{g(\alpha x)}{\alpha\sqrt\pi}  - \frac{g(\gamma x)}{\gamma \sqrt\pi}  + \frac{ \erfc(\alpha x)}{\beta\sqrt\pi} + \frac{e^{-(\gamma x)^2}}{\gamma \sqrt\pi} \erf\left(\frac{\alpha^2 x}{\sqrt{\alpha^2+\beta^2}}\right) - 4 x T\left(\sqrt{2}\gamma x, \frac{\alpha}{\beta}\right) \\
		g(x) &= e^{-x^2} - \sqrt\pi x \erfc(x)
	\end{align}
	We used $\gamma = \alpha\beta / \sqrt{\alpha^2 + \beta^2}$.

    
 \subsection{Integrals involving the Owen T function}
    \begin{align}
        \int_0^\infty dz\, T(\sqrt{2}\alpha z, a) &= \frac{1}{4\alpha \sqrt{\pi}} \frac{a}{\sqrt{1+a^2}} \\ 
        \int_x^\infty dz\,  T(\sqrt{2}\alpha z, b) dy &= \frac{1}{4\alpha \sqrt{\pi}} \left[\frac{a}{\sqrt{1+a^2}} \erfc\left(\alpha\sqrt{1+a^2}x\right) + e^{-(\alpha x)^2} \erf(\alpha a x)
		\right]
		-x T(\sqrt{2}ax, b) \\
        \int_0^\infty dz\, e^{-\alpha^2(z-x)^2} T(\sqrt{2}\beta z, b) &= \frac{\sqrt\pi}{8\alpha}\bigg[\frac{2}{\pi} \atan b \erfc(-\alpha x) - \erfc\left(-\frac{\alpha\beta}{\sqrt{\alpha^2+\beta^2}} x\right)\nonumber \\
        &+ 4 T\left(\frac{\alpha\beta\sqrt{2}}{\sqrt{\alpha^2+\beta^2}}x, \frac{b \alpha}{\sqrt{\alpha^2+(1+b^2)\beta^2}}\right) \\
        &+ 8 S\left(\frac{\alpha\beta\sqrt{2}}{\sqrt{\alpha^2+\beta^2}}x, \frac{b \alpha}{\sqrt{\alpha^2+(1+b^2)\beta^2}}, \frac{\sqrt{\alpha^2+\beta^2}}{b\beta}\right)\bigg] \nonumber \\
        \int_0^\infty dz\, T(\sqrt{2}\alpha z, a) \erfc(\beta z) &=
        \frac{a}{4\sqrt{1+a^2}\sqrt\pi \alpha} + \frac{\atan a}{2\pi^{3/2}\beta} -\frac{a}{2\sqrt{1+a^2}\pi^{3/2}\alpha}\atan\left(\frac{\beta}{\sqrt{1+a^2}\alpha}\right)\\
        &-\frac{\sqrt{\alpha^2+\beta^2}}{2\pi^{3/2}\alpha\beta}\atan\left(\frac{a\alpha}{\sqrt{\alpha^2+\beta^2}}\right) \nonumber 
    \end{align}
    \begin{align}
        \int_0^\infty dz\, T(\sqrt{2}\alpha z, a) T(\sqrt{2}\beta z, b) &= \frac{1}{8\pi^{3/2}}\bigg\{ \frac{a}{\sqrt{1+a^2}}\frac{1}{\alpha} \atan\left(\frac{b\sqrt{1+a^2}\alpha}{\sqrt{(1+a^2)\alpha^2 + (1+b^2)\beta^2}}\right)\nonumber\\
        &+\frac{b}{\sqrt{1+b^2}}\frac{1}{\beta} \atan\left(\frac{a\sqrt{1+b^2}\beta}{\sqrt{(1+a^2)\alpha^2 + (1+b^2)\beta^2}}\right) \\
        &-\frac{\sqrt{\alpha^2+\beta^2}}{\alpha\beta} \atan\left(\frac{ab\alpha\beta}{\sqrt{\alpha^2+\beta^2}\sqrt{(1+a^2)\alpha^2 + (1+b^2)\beta^2}}\right)
        \bigg\} \nonumber\\
        \int_0^\infty dz\, T(\sqrt{2}\alpha z, a) \erfc(\beta z) \erfc(\gamma z) &= \frac{1}{2\pi^{3/2}}\bigg[g_0(\alpha, \beta, \gamma, a) + g_0(\alpha, \gamma, \beta, a) \\ 
        &- \frac{\sqrt{\beta^2+\gamma^2}}{\beta\gamma} \atan\left(\frac{a\sqrt{\beta^2+\gamma^2}}{\sqrt{(1+a^2)\alpha^2 + \beta^2+\gamma^2}}\right)\bigg] 
    \end{align}
    \begin{multline}
        g_0(\alpha, \beta, \gamma, a) = \frac{1}{\alpha}\frac{a}{\sqrt{1+a^2}}\left[\atan\left(\frac{\alpha\sqrt{1+a^2}}{\beta}\right)-\atan\left(\frac{\alpha\sqrt{1+a^2}}{\beta}\frac{\gamma}{\sqrt{(1+a^2)\alpha^2 + \beta^2+\gamma^2}}\right)\right]
        + \frac{1}{\beta}\atan a \\
        + \frac{\sqrt{\alpha^2+\beta^2}}{\alpha\beta}\left[\atan\left(\frac{a\alpha}{\sqrt{\alpha^2+\beta^2}}\frac{\gamma}{\sqrt{(1+a^2)\alpha^2 + \beta^2+\gamma^2}}\right)-\atan\left(\frac{a\alpha}{\sqrt{\alpha^2+\beta^2}}\right)\right]
    \end{multline}
    
    \subsection{Integrals involving the Owen S function}
    \begin{align}
        \int_0^\infty dz\, S(-\sqrt{2}\alpha z, a, b) &= \frac{1}{4\pi^{3/2} \alpha}\left[\atan b - \frac{a}{\sqrt{1+a^2}}\atan\left(\frac{ab}{\sqrt{1+a^2}}\right) \right] \\
        \int_0^\infty dz\, S(\sqrt{2}\alpha z, a, b)\erfc(\beta z) &= \frac{1}{4 \pi ^{3/2}}\bigg\{
        \frac{1}{\beta }\atan\frac{a}{\sqrt{1+a^2(1+b^2)}} + \frac{\sqrt{a^2+\beta^2}}{\alpha\beta }\atan\left(b\sqrt\frac{\alpha^2+\beta^2}{(1+a^2(1+b^2))\alpha^2 + \beta^2}\right) \\
        & -\frac{1}{\alpha}\atan b + \frac{a}{\sqrt{1+a^2}}\frac{1}{\alpha}\left[\atan\frac{\sqrt{1+a^2}}{ab}\frac{\sqrt{(1+a^2(1+b^2))\alpha^2 + \beta^2}}{\beta}-\atan\frac{\sqrt{1+a^2}}{ab}\right]\bigg\} \nonumber
    \end{align}

    \subsection{Useful relations}
    \begin{gather}
        \erfc(x) + \erfc(-x) = 2 \\
        \erfc(x) \erfc(-x) = 8 T(\sqrt{2} x, 1) \\
        \atan x + \atan\frac{1}{x}  = \frac{\pi}{2} \\
        \atan\sqrt\frac{x y}{z(x+y+z)} + \atan\sqrt\frac{y z}{x(x+y+z)}  + \atan\sqrt\frac{x z}{y(x+y+z)}  = \frac{\pi}{2}
    \end{gather}

\bibliographystyle{apsrev4-2}

\end{document}